\documentclass[a4paper,12pt]{article}

\pdfoutput=1
\usepackage{graphicx}
\usepackage{epstopdf}
\usepackage[centertags]{amsmath}
\usepackage{amssymb}
\usepackage{amsthm}
\usepackage{amsfonts}
\usepackage{ccaption}
\usepackage[usenames]{color}
\usepackage{mathrsfs}
\usepackage[
      colorlinks=true,
      linkcolor=blue,  
      urlcolor=black,    
      filecolor=black,     
      citecolor = blue,
      pdfstartview=FitV,
      pdftitle={Dynamical black holes and expanding plasmas},
        pdfauthor={Pau Figueras, Veronika Hubeny, Mukund Rangamani, Simon Ross},
        pdfsubject={},
        pdfkeywords={fluid-gravity,bjorken flow, hydrodynamics, black holes, horizons},
        pdfpagemode=None,
        bookmarksopen=true    
      ]{hyperref}
       
\usepackage{rotating}

\makeatletter
\@addtoreset{equation}{section}

\makeatletter
\renewcommand\section{\@startsection {section}{1}{\z@}%
                                   {-3.5ex \@plus -1ex \@minus -.2ex}
                                   {2.3ex \@plus.2ex}%
                                   {\normalfont\large\bfseries}}
\renewcommand\subsection{\@startsection{subsection}{2}{\z@}%
                                     {-3.25ex\@plus -1ex \@minus -.2ex}%
                                     {1.5ex \@plus .2ex}%
                                     {\normalfont\bfseries}}

  \captionnamefont{\bfseries}
  \captiontitlefont{\small\sffamily}
  \captiondelim{: }
  \hangcaption

\def\baselinestretch{1.2}
\newcommand{\be}{\begin{equation}}
\newcommand{\ee}{\end{equation}}
\newcommand{\beq}{\begin{eqnarray}}
\newcommand{\eeq}{\end{eqnarray}}

\def\sec#1{\S\ref{#1}}
\def\fig#1{Fig.\,\ref{#1}}
\def\req#1{(\ref{#1})}
\def\App#1{Appendix \ref{#1}}

\def\p{\partial}

\def\CA{{\cal A}}
\def\CB{{\cal B}}
\def\CC{{ \cal C }}

\def\CF{{\cal F}}

\def\CN{{\cal N}}
\def\CO{{\cal O}}

\def\CS{{\cal S}}

\def\R{{\bf R}}
\def\Sp{{\bf S}}

\def\p{\partial}

\def\p{\partial}

\def\ord#1{\[ \, #1 \, \]}

\def\ord#1{\CO\left(#1\right)}

\def\AdS#1{AdS$_{#1}$}
\def\SAdS#1{Schwarzschild-AdS$_{#1}$}
\def\scri{\mathscr I}
\def\zs{z_s}
\def\ts{t_s}
\def\xs{x_s}

\title{{\bf \Large Dynamical black holes \& expanding plasmas}}

\author{\normalsize 
Pau Figueras$^a$\footnote{pau.figueras@durham.ac.uk}, \ 
Veronika E. Hubeny$^{a,b}$\footnote{veronika.hubeny@durham.ac.uk},\ 
Mukund Rangamani$^{a,b}$\footnote{mukund.rangamani@durham.ac.uk}, \ and 
Simon F. Ross$^a$\footnote{S.F.Ross@durham.ac.uk} \\
$^a$\small \sl  Centre for Particle Theory \& Department of
Mathematical Sciences,
\\[-1.5mm]
\small \sl Science Laboratories, South Road, Durham DH1 3LE, United Kingdom. \\
$^b$ \small \sl Kavli Institute for Theoretical Physics,
\\[-1.5mm]
\small \sl University of California, Santa Barbara, CA 93015, USA.
}

\begin{document}

\setlength{\baselineskip}{16pt}
\begin{titlepage}
\maketitle
\begin{picture}(0,0)(0,0)
\put(340, 270){DCPT-09/13} 
\put(340, 252){NSF-KITP-09-20}
\end{picture}
\vspace{-36pt}

\begin{abstract}
  We analyse the global structure of time-dependent geometries dual to expanding 
  plasmas, considering two examples: the boost invariant Bjorken flow, and the conformal soliton flow. While the geometry dual to the Bjorken  flow is constructed in a perturbation expansion at late proper time, the conformal soliton flow has an exact dual  (which corresponds to a Poincar\'e patch of Schwarzschild-AdS).  In particular, we discuss the position and  area of event and apparent horizons in the two geometries.  The
  conformal soliton geometry offers a sharp distinction between event
  and apparent horizon; whereas the area of the event horizon
  diverges, that of the apparent horizon stays finite and constant.
  This suggests that the entropy of the corresponding CFT state is
  related to the apparent horizon rather than the event horizon.
 \end{abstract}
\thispagestyle{empty}
\setcounter{page}{0}
\end{titlepage}

\renewcommand{\baselinestretch}{1.4}  
\renewcommand{\thefootnote}{\arabic{footnote}}


\tableofcontents

\section{Introduction}
\label{intro}

The AdS/CFT correspondence has over the years played an invaluable
role in providing insight into the dynamics of strongly coupled gauge
theories. An important application of the correspondence has been to
understand the holographic description of hydrodynamic properties of
field theories. This can be used to understand qualitative features of
the Quark-Gluon plasma (QGP) produced in heavy ion collisions. Current
theoretical understanding of this system is that subsequent to rapid
thermalization, the system evolves as an almost ideal fluid, expanding
rapidly away from the central collision region. The evolution in this
regime has been well described by the so called Bjorken flow
\cite{Bjorken:1982qr}.

The first step in understanding the holographic dual of the Bjorken
flow was taken in the seminal works \cite{Janik:2005zt,Janik:2006gp},
where the spacetime dual to the Bjorken flow in $\CN=4$ Super-Yang
Mills was constructed as a perturbation expansion at late times (see
\cite{Heller:2008fg} for recent review summarising the development of
the holographic description of the boost invariant plasma). This
geometry models the early (but post-thermalization) stages of the
expanding quark gluon plasma.

More generally, given any interacting quantum field theory, one can
study its hydrodynamic regime.  A natural question in the context of
the AdS/CFT correspondence is whether this regime admits a holographic
dual spacetime. This was answered in the affirmative in
\cite{Bhattacharyya:2008jc}, where the authors proposed the {\em
  fluid-gravity correspondence} relating the dynamics of the field
theory fluid to gravitational dynamics of black holes in an
asymptotically AdS spacetime. The fluid-gravity correspondence
generalizes the extensive discussion of hydrodynamics of field
theories with gravitational duals\footnote{For a review and references to earlier works on hydrodynamic aspects of $\CN =4$ Super Yang-Mills see  \cite{Son:2007vk}. Extensions of the fluid-gravity correspondence to include forcing and charge transport have been considered in \cite{VanRaamsdonk:2008fp,Bhattacharyya:2008ji,Haack:2008cp,Bhattacharyya:2008mz,Erdmenger:2008rm,Banerjee:2008th,Haack:2008xx,Hansen:2008tq,Caldarelli:2008ze}.}
and explicitly constructs spacetime
geometries dual to fluid flows in the hydrodynamic regime; that is,
for cases where the fluid remains in local thermal equilibrium. This
provides a useful relation between the dynamics of strongly coupled
systems in the long-wavelength regime and corresponding asymptotically
AdS black hole geometries. It provides a new approach to the
calculation of properties such as transport coefficients of the field
theory fluid.

In general, the flow of a viscous fluid, which involves dissipation, necessarily leads to 
entropy production. This of course is a simple consequence of the second law of thermodynamics.  In the geometric description of the fluid flow one can ask what this phenomenon of entropy production corresponds to.  An important ingredient in the fluid-gravity correspondence was the identification of the global event horizon in the bulk spacetime, which turned out to provide a simple geometric construction for a Boltzmann H-function in the bulk geometry. It was shown in~\cite{Bhattacharyya:2008xc} that the event horizon in the spacetimes dual to non-linear fluid flows could be determined
essentially locally despite the teleological nature of event horizons. This was achieved
by assuming slow temporal variations, as well as that the geometry will settle down to a stationary
configuration at late times (which is of course natural from the fluid
dynamical point of view, as one expects the dissipative effects of
viscosity etc., to cause the fluid motion to slow down asymptotically
and the system to achieve global equilibrium). The location of the
event horizon is then given by a perturbation around this final
equilibrium position. The perturbed position of the horizon can be
determined order by order in the derivative expansion. The area-form of this event horizon when pulled back to the boundary was shown to lead to a natural local entropy current with non-negative divergence as required by the second law.

Our main aim in this paper is to extend this work to determine the
location of the horizons in certain time-dependent geometries that do not
settle down to stationary finite-temperature solutions at late
times. Our interest in this question originally arises from the
Bjorken flow (BF) geometry, but we will also consider the conformal
soliton (CS) geometry, which provides a simpler example with stronger
time dependence.

For the Bjorken flow, explicitly constructing the event horizon will
allow us to confirm the regularity of the bulk geometry. In
\cite{Janik:2005zt,Janik:2006gp}, the geometry was constructed as a
perturbation expansion in the boundary time coordinate which is valid
at late times. By demanding regularity of the solution at leading
orders, the authors were able to derive the transport properties of
the plasma, most notably the shear-viscosity $\eta$ which saturates
the famous bound $\eta/s \ge 1/4\pi$ \cite{Kovtun:2004de}. The study
of the gravitational dual at higher orders was undertaken to derive
the relaxation time of the plasma in
\cite{Heller:2007qt,Baier:2007ix}. However, the regularity of the dual
spacetime was brought into question as subleading singularities were
encountered \cite{Heller:2007qt, Benincasa:2007tp}. This issue was
addressed recently in \cite{Heller:2008mb,Kinoshita:2008dq} where the
authors used the framework of the fluid-gravity correspondence
\cite{Bhattacharyya:2008jc} to argue that the spacetime was indeed
regular. We will revisit this analysis and confirm regularity by explicitly constructing the global event
horizon for these geometries. Previously, \cite{Kinoshita:2008dq}
found the apparent horizon in the BF geometry, as an approximation to
the event horizon; here we confirm that the actual event horizon indeed
closely tracks the location of the apparent horizon.\footnote{
Here we will take  ``apparent horizon" to mean the full co-dimension 1 surface in the spacetime rather than just a co-dimension 2 slice of that surface; please see the Note added in v2 at the end of Discussion for a clarification of the quasilocal horizon jargon.  We thank Roberto Emparan for valuable discussions on these issues.} Since the Bjorken
flow does not settle down to a finite-temperature stationary state, we
determine the event horizon by explicitly constructing the boundary of
the past of the future null infinity $\scri^+$.\footnote{The future
  null infinity $\scri^+$ corresponds to `endpoints' of
  future-directed null geodesics and is timelike for asymptotically
  AdS spacetimes.  } Curiously, we find that the apparent horizon lies
outside the event horizon at the leading order in the perturbation
expansion.  This simply reflects the fact that the leading order
metric violates energy conditions.  At first order the event horizon
overtakes the apparent horizon and the situation becomes more
conventional.

The conformal soliton geometry \cite{Friess:2006kw} provides a simpler
example, where we can gain intuition about more general fluid
flows. It is simply a patch of the well-known \SAdS{} black hole, so
the explicit metric is known exactly, and admits a high degree of
symmetry.  Nevertheless, if we work in a coordinate system
corresponding to considering the field theory on flat space $\R^{3,1}$ rather than the Einstein static universe $\Sp^3\times \R^1$, taking a `Poincar\'e patch' of the \SAdS{} black
hole, the time translation symmetry is no longer manifest, and the
solution looks highly dynamical.  Pictorially, it corresponds to a
black hole entering through the past Poincar\'e horizon and exiting
through the future Poincar\'e horizon, with its closest approach to
the boundary occurring at $t=0$ (Poincar\'e time).  In the boundary CFT, this describes
a finite energy lump which collapses and re-expands in a
time-symmetric fashion. Here the hydrodynamic approximation is not
valid at all times, but because this fluid flow is conformal to a
stationary fluid on the Einstein static universe, the stress tensor is
shear-free; that is, there is no dissipation in this fluid flow.

The horizons in this case are more interesting. Naively one might
expect that since we are just performing a coordinate transformation
on a given solution, any geometrical feature, such as the location of
the event horizon, remains invariant under such a transformation.  In
other words, we might expect that the event horizon of the Poincar\'e
patch of \SAdS{} black hole coincides with the event horizon of the
global \SAdS{}.  However, as we argue below, this is not the case,
because our coordinate patch now includes only part of the future
infinity of the global \SAdS{}. As a result, the actual event horizon
for the conformal soliton lies outside the global event horizon of
\SAdS{}. Indeed, we discover that the area of the CS event horizon
diverges at late times. 

This surprising result leads to a puzzle: if we associate the entropy
of the corresponding CFT conformal soliton state to the area of the CS
event horizon, as is usually assumed to be the case, then we find that
this entropy likewise diverges at late times.  But the conformal
soliton describes a shear-free flow, with no entropy production
whatsoever.  Said differently, the conformal transformation from the
CFT on $\Sp^3 \times \R^1$ to the CFT on $\R^{3,1}$ should
leave the entropy invariant.  But the former describes a perfect fluid
in global thermodynamic equilibrium: its entropy is finite and constant in time. 

In fact, it has been argued previously in several different contexts
\cite{Hubeny:2007xt,Chesler:2008hg} that it may be more appropriate to
associate the entropy of the CFT configuration to the area of the
apparent horizon rather than the event horizon in the bulk dual.  We
will show that the apparent horizon in this Poincar\'e slicing still
coincides with the global \SAdS{} event horizon, whose area is indeed
constant.  Thus, this is a case where the event horizon and the
apparent horizon are very different. The CS geometry therefore provides a
good testing ground for studying the distinction between the event and
apparent horizons and the role they play for the associated CFT dual.
We see that in this case the CFT entropy is clearly more naturally associated with the
latter rather than the former.

This might also seem enigmatic, as it was argued
in~\cite{Bhattacharyya:2008xc} that the apparent horizon and the event
horizon track each other closely in the hydrodynamic regime. Again,
the essential difference between the cases we consider here
and~\cite{Bhattacharyya:2008xc} is that in that general
analysis, it was assumed that the geometry would settle down at late
times to a stationary finite-temperature black hole. Neither of the
geometries we consider have this property. For the Bjorken flow, we
find qualitatively similar results, in that the apparent horizon and
event horizon nevertheless track each other closely. But for the
conformal soliton, the late time boundary conditions force the apparent
horizon and the event horizon to behave very differently.

In the next section, we consider the Bjorken flow, summarizing
previous work and determining the location of the event horizon. In
\sec{confsol}, we consider the horizons in the conformal
soliton, focusing on the three-dimensional case, where the
calculations are simplest. We conclude in \sec{disc} with a
discussion of the lessons of these examples and open problems for the
future.  The Appendices collect generalizations and some of the more technical arguments.

\section{Boost invariant flow}
\label{bjorken}

As described in the Introduction, the Bjorken flow (BF) plays a
central role in understanding the post-thermalization evolution of the
QGP produced in heavy-ion collisions. The basic physical picture
developed in \cite{Bjorken:1982qr} is that in the central rapidity
region of ultra-relativistic collisions of heavy ions, assuming local
thermal equilibrium, one can model the flow of the plasma via
quasi-ideal hydrodynamics. In the hydrodynamic description, it is
assumed that the fluid evolution respects the boost symmetry along the
collision axis. This implies a boost invariant expansion of the fluid,
consistent with the observed distribution of the particles in the
collision process.

\subsection{Bjorken hydrodynamics}
\label{bjprelim}

To understand the hydrodynamics in the BF, consider Minksowski
spacetime $\R^{3,1}$ written in Milne-type coordinates which respect
boost invariance in a $\R^{1,1}$ subspace, i.e.,
\begin{equation}
ds^2 = -d\tau^2 + \tau^2 \, dy^2 + dx_\perp^2 \ . 
\label{milne}
\end{equation}	
The coordinates $\tau$ and $y$ measure the proper time and rapidity in
the longitudinal direction respectively, and $x_\perp$ collectively
denotes the transverse directions. For a conformally invariant fluid,
the equations of motion of hydrodynamics, viz., energy momentum
conservation and tracelessness of stress tensor,
\begin{equation}
\nabla_\mu T^{\mu \nu } =0 \ , \qquad T_\mu^{\ \mu} =0\ ,
\label{fluideqns}
\end{equation}	
can be shown to constrain the dynamics to be derivable from a single
function $\varepsilon(\tau)$ which is conveniently taken to be the
energy density \cite{Janik:2005zt}. For an ideal conformal fluid, the
equations of motion lead to a power law fall-off for the energy
density and temperature
\begin{equation}
\varepsilon(\tau) = \frac{\varepsilon_0}{\tau^{\frac{4}{3}}} \ ,
\qquad  T \sim \tau^{-\frac{1}{3}} \ , 
\label{idealbf}
\end{equation}	
with the entropy per unit rapidity remaining constant. It is clear
from \req{idealbf} that there is a divergent amount of energy density
localized on the forward light-cone, $\tau =0$. Nevertheless, as the
fluid expands the energy density diffuses throughout the forward
light-cone. At late times the ideal hydrodynamic description becomes
more and more accurate in the interior of the light-cone in
$\R^{1,1}$.

One can in fact go beyond the ideal fluid description of the BF and
argue that an expansion in powers of $\tau^{-\frac{2}{3}}$ corresponds
to the derivative expansion in the fluid dynamics. Recall that the
hydrodynamic description can be thought of as an IR effective field
theory, valid at long wavelengths, for any interacting system that
achieves local thermal equilibrium. Given this, one can explore the
dissipative corrections to fluid dynamics by studying the system in a
perturbation expansion at large proper time $\tau$. This was carried
out in \cite{Janik:2006gp} to include viscous corrections. To derive
the transport properties of the plasma, the authors examined the
gravitational dual of the flow in the context of the AdS/CFT
correspondence.

\subsection{The gravity dual to Bjorken flow}
\label{bjgrav}

In \cite{Janik:2005zt,Janik:2006gp}, the authors also constructed the
bulk geometry dual to the given fluid dynamical evolution using the
AdS/CFT correspondence.\footnote{The gravity dual to Bjorken flow in 1+1 dimensions was discussed in \cite{Kajantie:2007bn}. Note that there isn't a hydrodynamic limit in 1+1 dimensions for conformal fluids. This is reflected in the bulk by the solutions being just the BTZ black hole written in a different coordinate chart. In \cite{Kajantie:2008hh} the dual spacetime to a spherically symmetric boost invariant flow was constructed
} The bulk metric was written in a
Fefferman-Graham type coordinate chart, and Einstein's equations were
solved to the desired order in a power series in $\tau^{-2/3}$ at late
times.  To be precise, consider an ansatz for the spacetime metric:
\begin{equation}
ds^2 = \frac{1}{z^2}\,\left( -e^{\alpha(\tau,z)} \, d\tau^2 + \tau^2 \, e^{\beta(\tau,z)}\, dy^2 + e^{\gamma(\tau,z)}\, dx_\perp^2 \right) + \frac{dz^2}{z^2} \ .
\label{fganz}
\end{equation}	
The Fefferman-Graham expansion involves solving Einstein's equations for the functions $\alpha$, $\beta$ and $\gamma$ perturbatively  in  the region $z \ll 1$, subject to asymptotic AdS boundary conditions, i.e., one expands the functions as\footnote{Recall that in the Fefferman-Graham coordinates used in \req{fganz} the boundary of the spacetime is at $z = 0$. We have also for brevity ignored $\log z$ terms which appear in even spacetime dimensions.}
\begin{equation}
\alpha(\tau,z) = \sum_{n=0} \, \alpha_n(\tau) \, z^{4 + 2n} \ ,
\label{}
\end{equation}	
and similarly for $\beta$ and $\gamma$.
This perturbatively constructed solution, it was argued, could be re-expressed in terms of a  scaling variable
\begin{equation}
v= \frac{z}{\tau^{\frac{1}{3}}} \ ,
\label{}
\end{equation}	
which then allows one to work at late proper times $\tau \gg 1$. This analysis leads to the above mentioned perturbation expansion in $\tau^{-2/3}$. For details we refer the reader to 
\cite{Janik:2005zt}.

By demanding that the spacetime thus constructed be regular, it was shown that the
shear-viscosity of the plasma saturates the universal lower bound
$\eta/s = 1/4 \pi$. To demand regularity, the authors looked at the
first non-trivial curvature invariant, the Kretschmann scalar, $R_{\mu
  \nu \rho\sigma}\, R^{\mu \nu \rho \sigma}$, expanded in a power
series in $\tau^{-2/3}$. Of course, well behavedness of a single
curvature invariant by itself does not guarantee that the spacetime is
completely regular, but this was sufficient to fix the transport
coefficients.

In \cite{Heller:2007qt,Benincasa:2007tp} this geometry was examined at
higher orders, and it was found that the spacetime appears to be
singular (at the third order). The appearance of this singularity
would seem contrary to the general analysis of the fluid-gravity
correspondence in \cite{Bhattacharyya:2008jc}, where black hole
solutions dual to arbitrary fluid flows in the boundary field theory
were constructed in a derivative expansion and conjectured to be
regular. However, the BF spacetime does not settle down to a
stationary finite-temperature black hole, so the demonstration of the
existence of a regular event horizon in~\cite{Bhattacharyya:2008xc}
does not apply in this case. 

In fact, the appearance of a singularity is associated with a poor
choice of coordinate system: in \cite{Janik:2005zt,Janik:2006gp} the
authors chose to work with the Fefferman-Graham coordinatization of
AdS, which was argued in \cite{Bhattacharyya:2008xc,
  Bhattacharyya:2008mz} to be problematic for discussing regularity
issues. Indeed, to avoid this subtlety, the original construction of
gravitational duals of fluid flows was done in ingoing
Eddington-Finkelstein type coordinates in
\cite{Bhattacharyya:2008jc}. In \cite{Heller:2008mb,Kinoshita:2008dq} (see also \cite{Kinoshita:2009dx})
the gravity dual to the BF was constructed in the
Eddington-Finkelstein coordinates, and it was argued that in these
coordinates the BF geometry is indeed regular.

Let us review the details of this construction: we will follow the
conventions of \cite{Kinoshita:2008dq} (except for using $\tau$ rather
than $\tau_+$ to denote the proper time coordinate). As we want the
bulk geometry to asymptote to \req{milne} on the boundary and
naturally adapt the coordinate chart to ingoing null geodesics, we
have a metric ansatz:
\begin{equation}
 ds^2=-r^2\,a\,d\tau^2+2\,d\tau \,dr+r^2\,\tau^2\,e^{2(b-c)}\,
 \bigg(1+\frac{1}{u\,\tau^{2/3}}\bigg)^2\,dy^2+r^2\,e^c\,dx_{\perp}^2\,, 
\label{boostmetric}
\end{equation}
where 
\begin{equation}
u\equiv r\,\tau^{1/3} \ ,
\label{udef}
\end{equation}	
and the functions $a$, $b$, $c$ depend on $u$ and $r$.

The idea is to use \eqref{boostmetric} as an ansatz and solve
Einstein's equations iteratively in a late time expansion, $\tau\to
\infty$, keeping $u$ fixed. In order to do so, one assumes that the
functions $a$, $b$ and $c$ can be expanded as
\begin{equation}
\begin{aligned}
 a(\tau,u)&=a_0(u)+a_1(u)\,\tau^{-2/3}+a_2(u)\,\tau^{-4/3}+a_3(u)\,\tau^{-2}
	+\ord{\tau^{-8/3}}\,, \\
b(\tau,u)&=b_0(u)+b_1(u)\,\tau^{-2/3}+b_2(u)\,\tau^{-4/3}+b_3(u)\,\tau^{-2}
	+\ord{\tau^{-8/3}}\,,   \\
c(\tau,u)&=c_0(u)+c_1(u)\,\tau^{-2/3}+c_2(u)\,\tau^{-4/3}+c_3(u)\,\tau^{-2}
	+\ord{\tau^{-8/3}}\,.
\end{aligned}
\label{abcexpn}
\end{equation}	
To solve the Einstein equations, one imposes the boundary conditions,
\begin{equation}
 a\big|_{u=\infty}=1\,,\qquad b\big|_{u=\infty}=0\,,\qquad c\big|_{u=\infty}=0\,,
\end{equation}
which ensure that the spacetime has boundary metric consistent with
the Bjorken flow, \req{milne}. We will denote the metric obtained by
this procedure at order $\ord{\tau^{-2k/3}}$ as $g^{(k)}$, so that
$g^{(k)}$ is specified completely by the functions $a_i(u)$, $b_i(u)$
and $c_i(u)$ for $i \le k$.

The Einstein's equations for gravity (with a negative cosmological
constant) were solved order by order in the late time expansion, and
the solutions depend only on a set of arbitrary constants. These
constants can be fully determined order by order by requiring that the
geometry be asymptotically AdS and that the Kretschmann scalar is
regular except at the origin $r=0$
\cite{Heller:2008mb,Kinoshita:2008dq}. This determines the choice of
transport coefficients along the lines of the original philosophy
espoused in \cite{Janik:2006gp}. This is different from the result of
\cite{Heller:2007qt,Benincasa:2007tp} because working in
Eddington-Finkelstein coordinates imposes regularity on the future
event horizon in the bulk geometry, which is the physically correct
condition. We should note however that the regularity of a particular curvature invariant is a necessary but not a sufficient condition for regularity. 

The rationale for the use of the Eddington--Finkelstein coordinates was given originally in \cite{Bhattacharyya:2008jc} (see also \cite{Bhattacharyya:2008mz}). To motivate this consider the hydrodynamic description of any interacting field theory; as explained earlier this makes sense so long as one achieves local thermal equilibrium. In the AdS/CFT context one expects each locally equilibriated domain in the field theory to have as gravity dual a stationary black hole solution.  These domains in the field theory extend into the bulk as ``tubes'' along ingoing null geodesics. In a sense, the construction of the gravity solution perturbatively in boundary derivatives corresponds to patching together these tubes (after all, this is what hydrodynamics achieves in the boundary description). Specifically, the tubes of relevance were argued to be centered along radially ingoing null geodesics, which in the coordinatization of \cite{Bhattacharyya:2008jc, Bhattacharyya:2008xc} are just $x^\mu  = constant$ with $x^\mu$ being the boundary coordinates. In the present case of the Bjorken flow, the Eddington-Finkelstein coordinates not only makes issues of regularity more transparent, but also provides a sensible coordinate chart to perform the late proper time expansion. Since there is no pathology in the coordinate chart in the zeroth order solution (the metric being completely regular there), it can then be shown that the higher order corrections in the late proper time expansion remain regular \cite{Heller:2008mb,Kinoshita:2008dq}.\footnote{ Essentially the distinction between the use of  Fefferman-Graham and the Eddington-Finkelstein coordinates may be traced to the trustworthiness of the late proper time expansion of various curvature invariants.}

At zeroth order, one obtains the spacetime with metric $g^{(0)}$:
\begin{equation}
 ds^2=-r^2\,\bigg(1-\frac{w^4}{u^4}\bigg)\,d\tau^2+2\,d\tau \,dr
	+r^2\,\tau^2\,\bigg(1+\frac{1}{u\,\tau^{2/3}}\bigg)\,dy^2+r^2\,dx_\perp^2\,,
	\label{bfzero}
\end{equation}
where $w$ is a constant, whose precise value will not play a role in
our discussion. This metric is consistent with the original derivation
given in \cite{Janik:2005zt}.  Note that this metric reduces to pure
AdS space for $w\to 0$. Naively it appears that \req{bfzero} is a
black hole metric with the location of the horizon being given by the
zero locus of $g^{(0)}_{\tau \tau}$, i.e.,
\begin{equation}
r(\tau) = \frac{w}{\tau^\frac{1}{3}} \ .
\label{naivehor}
\end{equation}	
For the metrics at higher orders, and a comprehensive discussion of
the derivation, we refer the reader to \cite{Kinoshita:2008dq}.

\subsection{Apparent horizon for the BF spacetime}
\label{bjregular}

To further bolster the claim that the spacetime dual to the BF
\req{boostmetric} is regular, in \cite{Kinoshita:2008dq} the apparent
horizon of the spacetime was determined explicitly up to second order
in the $\tau$ expansion. The presence of an apparent horizon implies
by virtue of the singularity theorems that the spacetime will evolve
into a singularity in the future. The idea of \cite{Kinoshita:2008dq}
was to argue that this apparent horizon must be enclosed by a global
event horizon, concealing the singularities from the asymptotic
region. This is a plausible argument, but to make it rigorous we would
need to check that the spacetime asymptopia is
complete.\footnote{Showing that the event horizon is outside the
  apparent horizon also requires that appropriate energy conditions
  are satisfied. Since the full geometry solves the vacuum Einstein
  equations with a cosmological constant, the null energy condition is of
  course satisfied. As we'll see below, however, if we work order by
  order in perturbation theory, at low orders the energy conditions
 are not be satisfied.} This requires a better understanding
of the global structure, which is the main focus of the present
work. We will explicitly construct the event horizon in this geometry
in the next section, and demonstrate that the spacetime is regular on
and outside the event horizon. Before turning to that however, it will
be useful to review the construction of the apparent horizon in
\cite{Kinoshita:2008dq}.

The apparent horizon is given by the null hypersurface for which the
expansion of the outgoing null geodesics vanishes. For the metric
\eqref{boostmetric}, the vectors tangent to the ingoing and outgoing radial
null geodesics are given by
\begin{equation}
l_{-}^a = -\left(\frac{\partial}{\partial r}\right)^{\!a}\,,\qquad l_{+}^{a} = \left(\frac{\partial}{\partial \tau}\right)^{\!a}+\frac{r^2\,a}{2}\left(\frac{\partial}{\partial r}\right)^{\!a} \,,
\end{equation} 
up to some overall normalisation factors that are irrelevant for this
calculation. Note that we are considering congruences that emanate normal to the co-dimension two spacelike surface in the spacetime and are exploiting the symmetries of the geometry to restrict our attention to normals  pointing in the radial direction.
Then the expansions $\theta_{\pm}$ are defined as
\begin{equation}
 \theta_{\pm}=\mathcal L_{\pm}\ln\mu\,,
\end{equation}
where $\mathcal L_{\pm}$ denotes the Lie derivative along $l^a_\pm$
and $\mu$ is the volume of the null hypersurface,\footnote{This
  exploits crucially the fact that the vectors
  $\left(\frac{\partial}{\partial y}\right)^a$ and
  $\left(\frac{\partial}{\partial x_\perp}\right)^a$ are Killing in
  the geometry \req{boostmetric}.}
\begin{equation}
 \mu=r^3\,\tau\,e^b\left(1+\frac{1}{\tau\, r}\right)\,.
\end{equation}

The quantity $\Theta=\theta_+\theta_-$ is an invariant,\footnote{More
  precisely, it is invariant under reparametrisations of the scalars
  that define the null hypersurface.} and hence the location of the
apparent horizon can be found by solving
\begin{equation}
 \Theta = 0  \label{apphor}
\end{equation}
for $r(\tau)$. Since the geometry is only known
in a perturbative expansion in $\tau$ for large proper time, the
location of the apparent horizon will also be determined in the power
series. Writing
\begin{equation}
r_A(\tau) = r_{A0} \, \tau^{-\frac{1}{3}}  + r_{A1} \, \tau^{-1} + r_{A2} \, \tau^{-\frac{5}{3}} + \ord{\tau^{-\frac{7}{3}}} \ ,
\label{raexp}
\end{equation}	
substituting this expansion into \eqref{apphor}, and using the
previously determined functions $a$ and $b$, we can find $r(\tau)$
order by order in $\tau^{-2/3}$. In \cite{Kinoshita:2008dq} this
procedure was carried out up to the second order for the metric
$g^{(2)}$, with the result:\footnote{In what follows use the notation $r^{(k)}_{Ai}$ and $r^{(k)}_{Ei}$ to denote the coefficients in the expansion of the apparent and event horizon
for the metric $g^{(k)}$.}

\begin{equation}
 r^{(2)}_{A0}=w\,,\qquad r^{(2)}_{A1}=-\frac{1}{2}\,,\qquad r^{(2)}_{A2}=\frac{8+3\pi-4\,\ln 2}{72\,w}\,. 
\label{locapp}
\end{equation}
It is not surprising that the zeroth order location of the apparent
horizon coincides with the naive horizon \req{naivehor}.

\subsection{Event horizon for the BF spacetime}
\label{bjevent}

As discussed earlier, to convincingly demonstrate the regularity of
the spacetime \req{boostmetric}, we have to show that the spacetime
has a well behaved global event horizon. This is defined as the
boundary of the past of future infinity $\scri^+$. It is by definition a
null surface and furthermore since it is the boundary of a causal set,
is generated by null geodesics. In the cases where the solution
settles down to a stationary configuration asymptotically, we know the
position of the horizon at late times, and can evolve back using the
geodesic equation to determine the location of the event horizon. In
the present case however, \req{boostmetric} does not appear to settle
down to a known stationary configuration. We will therefore determine
the location of the horizon directly, by studying the geodesic motion
on the spacetime \req{boostmetric} and determining which points cannot
send signals to infinity. The analysis is simplified by the fact that
\eqref{boostmetric} is co-homogeneity two, so the problem reduces to
studying geodesic motion in the $(r,\tau)$ plane. 

Since one has three Killing fields $\left(\frac{\partial}{\partial
    y}\right)^{\! a}$ and $\left(\frac{\partial}{\partial
    x_\perp}\right)^{\! a}$ the location of the horizon is simply
given by a curve $r(\tau)$. The null geodesic equation reduces to
\begin{equation}
 \frac{d}{d\tau}\,r(\tau)=\frac{1}{2}\,r(\tau)^2\,a(r(\tau),\tau)\,.
\label{nullgeods}
\end{equation}
The event horizon is the outermost solution of this equation which
does not reach $r= \infty$ at finite $\tau$.  At late times, $r(\tau)$
for the event horizon can be shown to admit an expansion in
$\tau^{-2/3}$ of the form
\begin{equation}
 r_E(\tau)=r_{E0}\,\tau^{-1/3}+r_{E1}\,\tau^{-1}+r_{E2}\,\tau^{-5/3}+\ord{\tau^{-7/3}}\,,
\label{rexp}
\end{equation}
where the $r_{Ei}$'s are some yet to be determined constants. These
constants can be found by solving \eqref{nullgeods} order by order
using the previously determined expansion for $a(\tau,u)$ at the same
order. Using the metric $g^{(2)}$ quoted in \cite{Kinoshita:2008dq} we
find that
\begin{equation}
 r^{(2)}_{E0}=w\,,\qquad r^{(2)}_{E1}=-\frac{1}{2}\,,\qquad r^{(2)}_{E2}=\frac{12+3\pi-4\,\ln 2}{72\,w}\,,
\label{locevent}
\end{equation}
This gives the location of the event horizon up to second order in the
late time expansion.

Comparing, \req{locevent} with \req{locapp}, we see that the event
horizon indeed lies outside the apparent horizon at this
order. Furthermore, the spacetime metric \eqref{boostmetric} is
regular on and outside the event horizon: the singularity at $r=0$ is
cloaked by the event horizon. We have thus demonstrated that the
spacetime at second order in the perturbation expansion, with metric
$g^{(2)}$, is indeed regular.

\paragraph{A curiosity at leading order:} The location of the event
horizon differs from the location of the apparent horizon derived in
\cite{Kinoshita:2008dq}  at second order in the expansion in
$\tau^{-2/3}$. Of course, we need to work with the second order metric
$g^{(2)}$ to study the position of the horizon to this order. However,
it is worth remarking on a curious behaviour which is seen if we work
with the metric at the zeroth order, and ask about the difference
between the locations of the apparent and event horizons.

\begin{figure}[t]
\begin{center}
\includegraphics[width=6in]{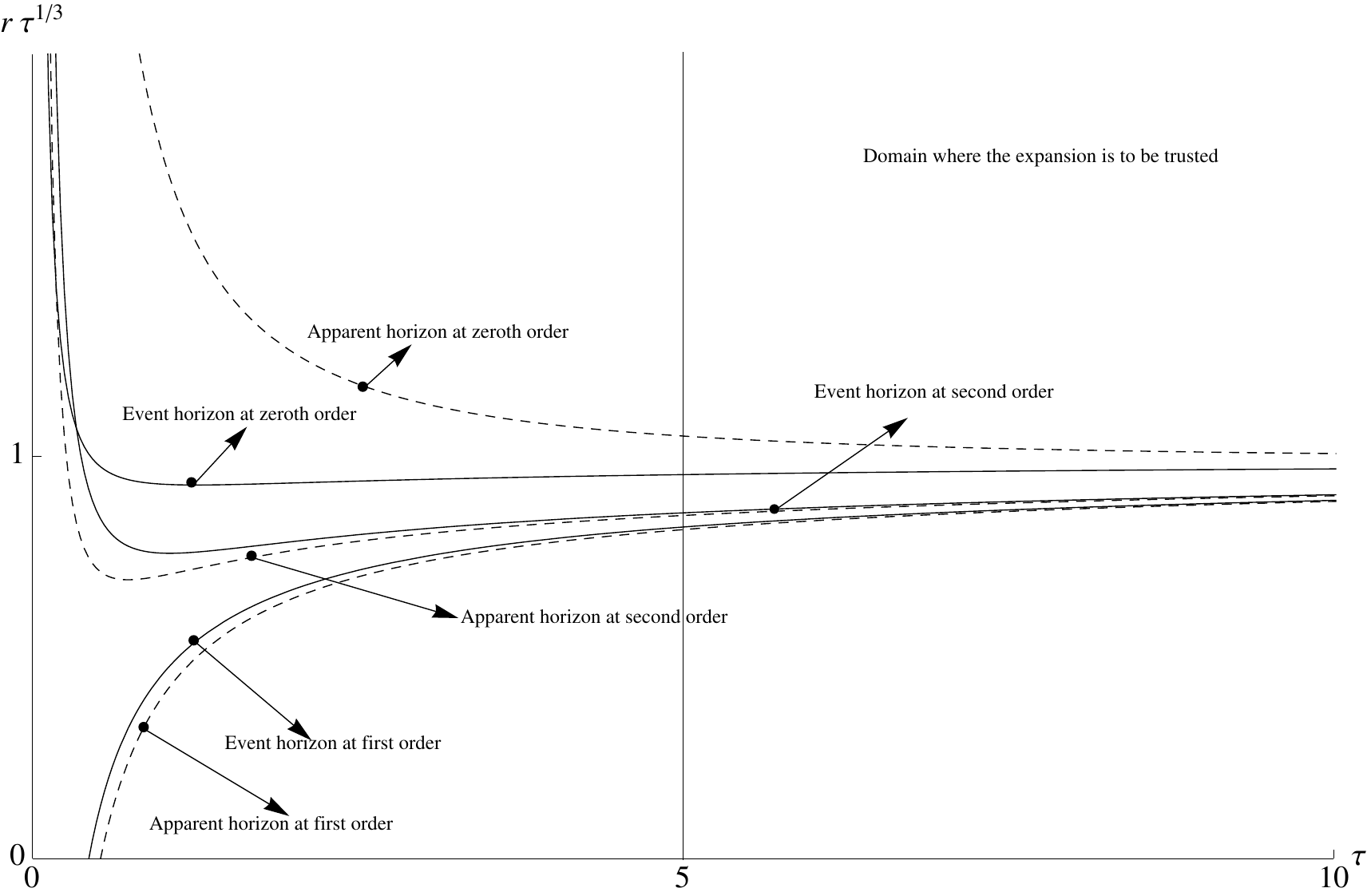}
  \caption{Illustration of the horizons for the Bjorken flow metrics $g^{(k)}$ at various orders in the perturbation expansion. The event horizons are the solid curves while the apparent horizons are the dashed curves. The locations of the horizons of course should only be trusted at late times as indicated in the figure. }
\label{fig:bfhors}
\end{center}
\end{figure}

It is clear that we should trust the coefficient $r_k$ in the
expansion for the location of the apparent/event horizon in
\req{raexp}, \req{rexp} only upon using the metric $g^{(k)}$. This is
because the metric $g^{(k)}$ only satisfies Einstein's equations to
$\ord{\tau^{-2k/3}}$.  If we consider the metric $g^{(0)}$, we can
thus only trust $r_{A0}$ and $r_{E0}$ in the expansions for the
apparent horizon, \req{locapp}, and the event horizon,
\req{locevent}. The event and apparent horizons lie on top of each
other at this order. Nevertheless, given the metric at some order we
can ignore the fact that it doesn't satisfy the appropriate field
equations and treat the residue as some effective energy-momentum
tensor required to support the geometry. One can then study the metric
at any given order as a spacetime in its own right and ask for the
locations of the apparent and event horizons at higher orders in the
$\tau$ expansion. For $g^{(0)}$ we find
\begin{eqnarray}
r^{(0)}_{A0} &=& w \ , \qquad r^{(0)}_{A1} = -\frac{1}{6} \ , \qquad r^{(0)}_{A2} = \frac{11}{12\, w}, \nonumber \\
r^{(0)}_{E0} &=& w \ , \qquad r^{(0)}_{E1} = -\frac{1}{6}\ , \qquad  r^{(0)}_{E2} = \frac{7}{72\, w} \ .
\label{zeroae}
\end{eqnarray}	
We see that the apparent horizon for the artificial metric $g^{(0)}$ lies outside the event horizon! This contradicts the expected behaviour (seen in $g^{(2)}$) that the
apparent horizon should lie behind the event horizon. We illustrate the location of the horizons at various orders in \fig{fig:bfhors}. 

The explanation for the apparent horizon lying outside the event horizon is simple: the geometry $g^{(0)}$ fails to be a solution of the vacuum equations beyond the leading order in perturbation theory, and the required stress tensor source violates the energy
conditions. This can be checked by computing $T^{\text{bulk}}_{\mu \nu} = R_{\mu\nu} - \frac{1}{2} \, R \, g_{\mu \nu} - 6 \, g_{\mu \nu}$ and seeing that the null energy condition is violated for \req{bfzero}. 

We have thus seen that the BF geometry is a regular black hole spacetime. Despite the dual fluid flow not quite settling down at late times as required for the analysis of \cite{Bhattacharyya:2008xc}, the spacetime event horizon can nevertheless be inferred by appropriate ray tracing.

\section{The Conformal Soliton flow}
\label{confsol}

In this section we will consider the conformal soliton (CS) geometry discussed initially in 
\cite{Friess:2006kw} as our second example of a dynamical flow.  The CS spacetime is extremely simple -- it is just the global AdS black hole sliced in Poincar\'e coordinates. 

As explained in \sec{intro}, to obtain the CS$_{d+1}$ spacetime we take the global \SAdS{d+1} black hole, which is dual to a fluid in global thermal equilibrium
in the Einstein Static Universe $\Sp^{d-1} \times \R^1$, and consider it in
a `Poincar\'e patch'. From the dual field theory point of view, we are
making a conformal transformation to map the field theory on $\Sp^{d-1}
\times \R^1$ to the field theory on Minkowski space, $\R^{d-1,1}$. This
maps the stationary fluid on the Einstein Static Universe to a
time-dependent fluid configuration on Minkowski space. From the bulk
spacetime point of view, this corresponds to
considering only the portion of the null infinity $\scri^+$ of
global \SAdS{d+1} restricted to this Minkowski patch in the Einstein
Static Universe; we call this subregion on the boundary
$\scri^+_{CS}$.  The corresponding Poincar\'e patch in the bulk contains not only the region outside the black hole which is simply the portion of global \SAdS{d+1} visible from this portion of null infinity, but also a finite region inside the black hole.
The former is bounded by past and future Poincar\'e horizons, as in the description
of global AdS in Poincar\'e coordinates, whereas the latter covers a larger 
region, whose boundary we'll refer to as ``Poincare edge".

We will be interested in the global structure of the solution and will find the event horizon in \sec{cseh} and the apparent horizon in \sec{csah}. For simplicity, we will consider the situation in 
$2+1$ dimensions, i.e., concentrate on the BTZ black hole. This lower dimension example captures all of the essential features of the calculation and has the significant advantage of being algebraically simpler. We will comment on the extension to higher dimensions in \App{cshighd}.  Also, without loss of generality we will set the AdS radius to unity, which translates to measuring all lengths in AdS units.

\subsection{The BTZ spacetime as a conformal  soliton}
\label{btzconfsol}

We follow \cite{Friess:2006kw}  and describe a region in the bulk geometry of the \SAdS{d+1} spacetime  in Poincar\'e coordinates, applying a specific coordinate transformation, one that transforms global \AdS{d+1} into Poincar\'e AdS, to the black hole spacetime.  

The coordinate transformation between the global coordinates $\{\tau,r,\phi\}$ in which pure AdS$_3$ has metric
\begin{equation}
ds^2 = - (r^2 + 1) \,  d\tau^2 + \frac{dr^2}{r^2+1} + r^2\, d\phi^2 
\label{AdSglobal}
\end{equation}	
and the Poincar\'e coordinates $\{t,z,x\}$ in which the metric is 
\begin{equation}
ds^2 = \frac{-dt^2 + dz^2 + dx^2}{z^2}
\label{AdSPoinc}
\end{equation}	
can be written as\footnote{This transformation can be easily obtained by writing \req{AdSglobal}  and \req{AdSPoinc} in terms of embedding coordinates describing a
  hyperboloid in $\R^{2,2}$.
Note, however, that this is not the only coordinate transformation which implements the desired conformal transformation on the boundary; in fact, it is not even the simplest one.  In \App{csapphor} we use an algebraically simpler transformation, which has the same limiting relations on the boundary $r \to \infty$.
  }
\begin{equation}
z = \frac{1}{\sqrt{r^2 +1} \, \cos \tau + r\, \cos \phi} \ , \ \
t = \frac{\sqrt{r^2 +1} \, \sin \tau }{\sqrt{r^2 +1} \, \cos \tau + r\, \cos \phi} \ , \ \
x = \frac{ r \, \sin \phi}{\sqrt{r^2 +1} \, \cos \tau + r\, \cos \phi} \ .
\label{gtop}
\end{equation}	
When this transformation is applied to pure AdS spacetime,
the Poincar\'e edge at $z = \infty$ actually coincides with the  Poincar\'e horizon, which is the null surface bounding the causal wedge of $\scri_{CS}$.
In the global coordinates, this surface is given by the relation 
$\sqrt{r^2 +1} \, \cos \tau + r\, \cos \phi = 0$.  Note that this relation describes the Poincar\'e edge if we apply the coordinate transformation \req{gtop} to a more general asymptotically-AdS spacetime as well; but in the general case this Poincar\'e edge no longer coincides with the null Poincar\'e horizon.

\begin{figure}[h]
\begin{center}
\includegraphics[scale=0.53]{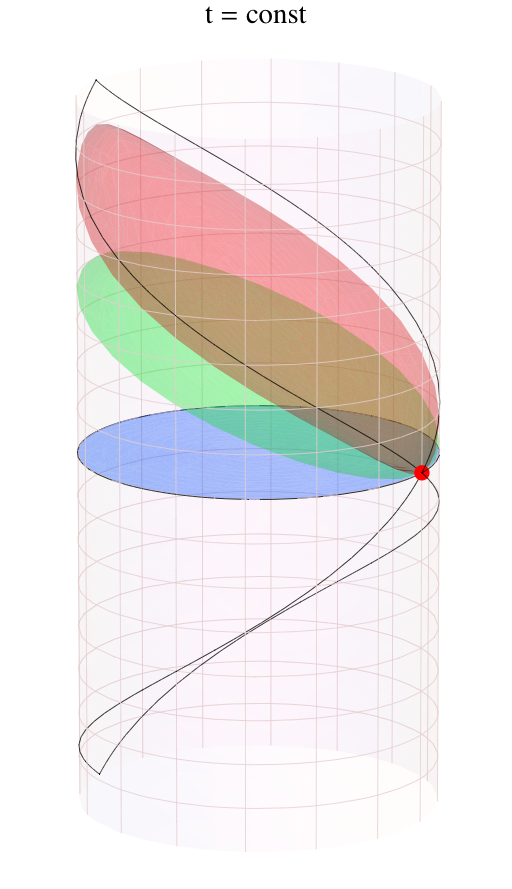}
\hspace{0.15cm}
\includegraphics[scale=0.53]{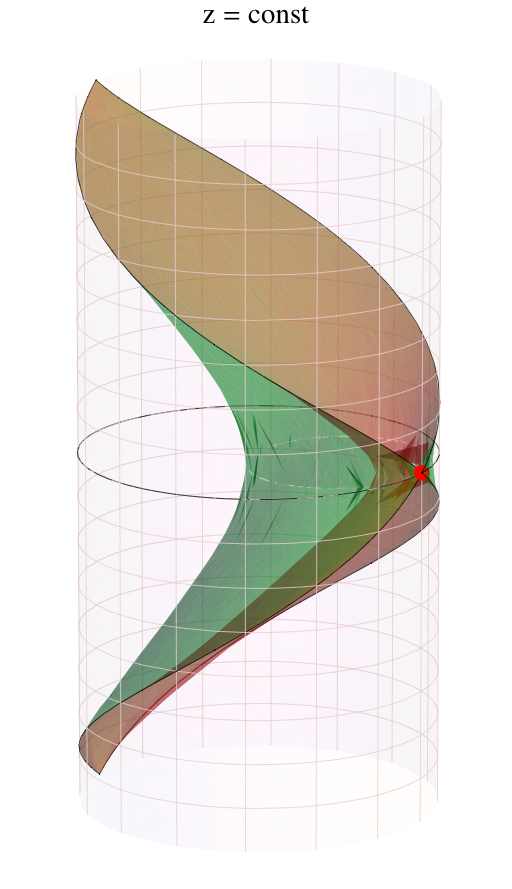}
\hspace{0.15cm}
\includegraphics[scale=0.53]{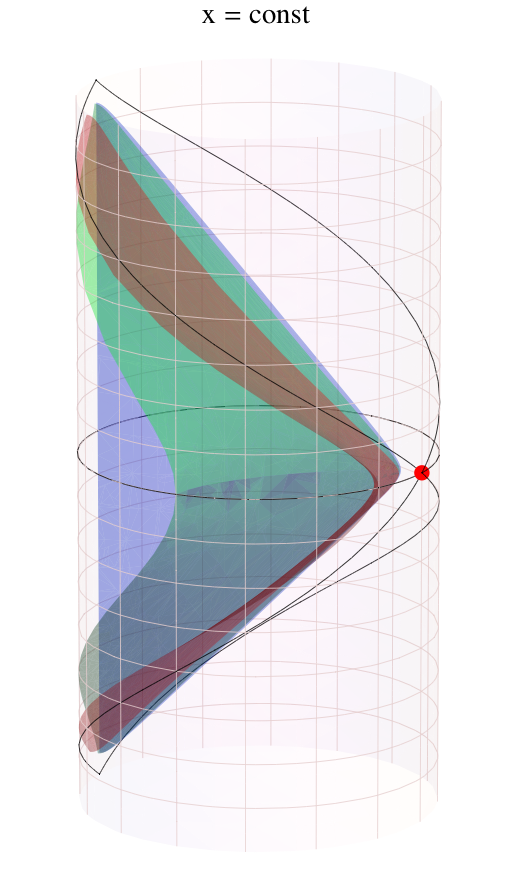}
  \caption{Illustration of Poincar\'e coordinates superposed on conformally compactified AdS.  The surfaces $t=0,1,5$ (left), $z=1,5$ (middle), and $x=0,1,5$ (right) are plotted, with colour-coding blue for 0, green for 1, and red for 5.  To guide the eye, surperposed are also the boundary of the $\scri_{CS}^+$ corresponding to $t= \pm \infty$ and the $t=0$ boundary slice (black curves). }		
\label{fig:AdScoords}
\end{center}
\end{figure}
\fig{fig:AdScoords} gives a plot of the constant Poincar\'e coordinates in the AdS spacetime.  The constant Poincar\'e time $t$ surfaces (left plot) are all pinned at the red reference point $i_{CS}^0$ (which in global coordinates corresponds to $\tau = 0, \phi = \pi$, and $r= \infty$), with the $t = \pm \infty$ surfaces coinciding with the  Poincar\'e edge.  The constant $z$ surfaces (middle plot) interpolate from the $\scri_{CS}^+$ at $z=0$ to the  Poincar\'e edge at $z = \infty$, with the surfaces having round cross-sections at $t=0$ slice, tangent to $i_{CS}^0$.  Finally, the constant $x$ slices (right plot) are pinned at  $i_{CS}^0$, and interpolate from a section of $\phi=0,\pi$ plane at $x=0$ to half of the Poincar\'e edge at $x = \pm \infty$.

Now, consider the BTZ spacetime with metric 
\begin{equation}
ds^{2} = -(r^2 -r_+^2 ) \, d\tau^2 + \frac{dr^2}{r^2 -r_+^2} + r^2\, d\phi^2 
\label{BTZ}
\end{equation}	
and perform the coordinate transformation \req{gtop} to obtain the
metric in Poincar\'e coordinates.  The resulting geometry is what we
call the CS spacetime.  We do not give the form of the bulk geometry in
these coordinates explicitly, as even in the simple BTZ case it is
rather messy and unilluminating, but it is clear that the metric will
appear time-dependent in these coordinates.\footnote{In fact, given that the 
coordinate transformation \req{gtop} involves all three coordinates, the resulting metric in the Poincar\'e coordinates has  no manifest symmetries.}

As required, the coordinate transformation \req{gtop} corresponds to a conformal transformation on the boundary of the AdS spacetime. It maps the Einstein Static Universe to Minkowski space. The conformal transformation can be inferred by restricting the transformation \req{gtop}  to the boundary ($r\to \infty$), where it results in the map:
\begin{equation}
t = \frac{\sin \tau}{\cos \tau + \cos \phi} \ , \qquad x = \frac{\sin \phi}{\cos \tau + \cos \phi} \ . 
\label{bdyconf}
\end{equation}	
It is easy to check that \req{bdyconf} maps the Lorentzian cylinder $\Sp^1 \times \R^1$ to $\R^{1,1}$,
\begin{equation}
ds^2 = -dt^2 + dx^2  = W^2 \, (-d\tau^2+ d\phi^2),  
\label{cynconf}
\end{equation}	
with 
\begin{equation}
W= \frac{1}{\cos \tau + \cos \phi} = \frac{1}{2}\, \sqrt{4\,t ^2 + (1+r^2 - t^2)^2}
\label{Wdef}
\end{equation}	

In the dual CFT description, the global \SAdS{} black hole corresponds to a static ideal fluid in thermal equilibrium. The field theory stress tensor transforms homogeneously under conformal transformations \cite{Wald:1984ai}, so applying the conformal transformation \eqref{bdyconf} will give a time-dependent fluid in  Minkowski space, but one which still describes an ideal fluid.  That is, this time-dependent fluid flow is free of dissipation. We give a brief review of the fluid description in \App{cshydro} and refer the reader to \cite{Friess:2006kw} for a comprehensive discussion in the \AdS{5} case.

\subsection{Event horizon for the CS spacetime}
\label{cseh}

Since we are restricting consideration to a subregion $\scri^+_{CS}$ of the full
future null infinity $\scri^+$ of the global \SAdS{} spacetime, the 
event horizon of the CS spacetime will be different from the event
horizon in global \SAdS{}. This event horizon will constitute the
boundary of the region of spacetime visible from $\scri^+_{CS}$, so it
is the proper analogue of the Poincar\'e horizon in pure
AdS. Intuitively, we expect it to interpolate smoothly between the
black hole horizon $r=r_+$ at very early times and the surface $z =
\infty$ close to the boundary, corresponding to our naive picture
of the CS spacetime as describing a black hole falling across the
Poincar\'e horizon.

A-priori, one might worry that from a gravitational viewpoint, consideration of $\scri^+_{CS}$ seems rather ad hoc, since we are by fiat restricting to a subset of the maximally extended spacetime's $\scri^+$. However, this is well-justified by the field theory. In the AdS/CFT correspondence one prescribes a conformal structure for the boundary. Then a  given boundary metric corresponds to a particular representative and one studies field theory on a background manifold with this prescribed metric. However, one is free to change the background on which the field theory lives. For the field theory on the Einstein Static Universe, one considers the global spacetime, whereas for the field theory on Minkowski space we are required to restrict attention to the Poincar\'e patch. For an observer in this boundary Minksowski space we have to define the horizon using $\scri^+_{CS}$. Our construction can be thought of as an observer dependent horizon in the spacetime, with the Minkowski observer being singled out by field theoretic considerations.

We could construct this horizon by working with the CS geometry in the
Poincar\'e coordinates, and finding the null surface with the
requisite late time behaviour, analogously to \sec{bjevent}.  However,
this is rather impractical. Instead, we can work with the metric
\req{BTZ} in the global coordinates, and look for the surface which
bounds the past of $\scri^+_{CS}$. In BTZ coordinates, we can take
$\scri^+_{CS}$ to be the connected region at $r=\infty$ containing
$\tau=0$ with $\cos \tau + \cos \phi \ge 0$. The CS event horizon is
then the boundary of the causal past of this region. Since all points
in $\scri^+_{CS}$ lie in the causal past of the boundary point
$(r=\infty, \tau = \pi, \phi=0)$, the problem of finding the CS event
horizon reduces to the problem of finding null geodesics ending on
this point.

Finding such null geodesics in \req{BTZ} is a straightforward
exercise.  The main subtlety arises from the fact that the null
geodesics ending on this boundary point caustic, so they do not form a
smooth surface.  Indeed, the presence of caustic points is typical for
an event horizon in a generic dynamical spacetime; here the caustic
locus is actually very simple, occurring only at $\phi = \pi$ (which
is identified with $\phi = -\pi$). A caustic at $\phi = \pi$ is
expected from the symmetry under $\phi \to -\phi$.  It is easy to show
that the event horizon is smooth for $\phi \ne \pm \pi$.  We now
proceed to determine this event horizon explicitly. 

The geodesic equations are
\begin{equation}
\dot{r}^2 = 1-\ell^2 + \frac{\ell^2\, r_+^2}{r^2} \ , \qquad
\dot{\tau} = \frac{1}{r^2 - r_+^2} \ , \qquad \dot{\phi} =
\frac{\ell}{r^2},
\label{BTZgeod}
\end{equation}	
where $\dot{} = \frac{d}{d\lambda}$ with $\lambda$ being the affine
parameter along the geodesic and $\ell$ is the conserved quantity
along each geodesic corresponding to angular momentum per energy.  We
can think of $\ell$ as specifying which geodesic we take and $\lambda$
as the position along that geodesic. Equations \req{BTZgeod} can be immediately
integrated to give
\begin{eqnarray}
r(\lambda,\ell)^2 &=& \left(1-\ell^2\right) \, \lambda^2 - \frac{\ell^2\,r_+^2 }{1-\ell^2}, \nonumber \\
\tau(\lambda,\ell) &=& \pi - \frac{1}{r_+} \, {\rm arccoth}\left(\frac{1-\ell^2}{r_+} \, \lambda\right) ,\nonumber \\
\phi(\lambda,\ell) &=&    -\frac{1}{r_+} \, {\rm arccoth}\left(\frac{1-\ell^2}{\ell\,r_+} \, \lambda\right), 
\label{gBTZlaml}
\end{eqnarray}	
where we have chosen the constants of integration so that the
geodesics have a future endpoint at $(r=\infty, \tau = \pi, \phi=0)$.
The relations \req{gBTZlaml} describe a 2-surface parameterized by
$\lambda$ and $\ell$ with $0 \le \ell < 1$ and $\lambda_{{\rm
min}}(\ell) \le \lambda < \infty$.  This 2-surface corresponds to the
CS event horizon.  Note that the geodesics with $\tanh(r_+ \pi)< \ell
<1$ terminate on the line of caustics at $\phi = \pi$, and
$\lambda_{{\rm min}}$ is determined by cutting off the surface when
the geodesics caustic. The caustic locus is obtained by solving
$\phi(\lambda, \ell) = \pi$, giving
\begin{equation}
r_c (\ell)  = -\frac{\ell}{\sqrt{1-\ell^2}} \, \frac{r_+}{\sinh (\pi
  \,r_+)}, \quad \tau_c(\ell) = \pi - \frac{1}{r_+} {\rm arccoth}
  (\ell \coth(\pi \,r_+ )). 
\label{btzcaus}
\end{equation}	
This generates the curve of caustics, described by a relation between
$\tau$ and $r$, with $\phi = \pi$, given by
\begin{equation}
\tau_c(r)  =\pi - \frac{1}{r_+} \, {\rm arctanh}\,\left(\frac{\sqrt{r^2\, \sinh^2(\pi\, r_+) + r_+^2}}{r\, \cosh(\pi \,r_+)}\right) .
\label{}
\end{equation}	
One important detail to note here is that the curve of caustics starts to exist only for $\ell$ larger than some minimum value $\ell_{\text{cmin}}$, where\footnote{
Note that in order to stay within the CS spacetime (in particular to the future of the past Poincar\'e edge), we need a stronger constraint: rather than bounding $\ell$ by $\tau \to - \infty$, we will bound $\ell$ by $t \to - \infty$, which provides a more stringent bound.
} $\tau \to -\infty$. In particular, 
\begin{equation}
\ell_{\text{cmin}} = \tanh(\pi\, r_+) \ ,
\label{lmin}
\end{equation}	
and $r_c(\ell_{\text{cmin}}) = r_+$. For large BTZ black holes $r_+ >1$, $\ell_{\text{cmin}}$ gets exponentially close to unity.

Instead of parameterising the horizon by $\lambda$ and $\ell$ as in \req{gBTZlaml}, it is in practice a bit simpler, though physically equivalent, to use
the fact that $r(\lambda)$ is a monotonic function, and think of the
event horizon as a surface parameterised by $r$ and $\ell$:
\begin{eqnarray}
\tau(r,\ell) &=& \pi - \frac{1}{r_+} \, {\rm arccoth} \left(\frac{\sqrt{(1-\ell^2) \, r^2 + \ell^2 \, r_+^2}}{r_+} \right), \nonumber \\
\phi(r,\ell) &=& -\frac{1}{r_+}\, {\rm arcsinh}\left(\frac{\ell \, r_+}{\sqrt{1-\ell^2}}\;\frac{1}{r}\right).
\label{gBTZrl}
\end{eqnarray}	

It is easy to confirm that the surface described by \req{gBTZlaml} or
\req{gBTZrl} is indeed a null surface.  For instance, the induced
metric on this 2-surface is simply
\begin{equation}
ds_{ind}^2 = \frac{d\ell^2}{(1-\ell^2)^2} ,
\label{hormet}
\end{equation}	
which is clearly degenerate, as required of the event horizon.

\begin{figure}[t]
\begin{center}
\includegraphics[scale=0.95]{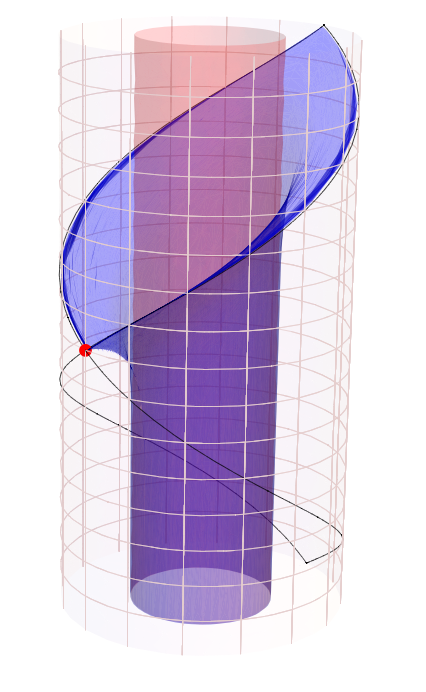}
\caption{Plot of the event horizon of the CS spacetime. For ease of visualization we have also plotted the location of the event horizon of the global \SAdS{} black hole.}
\label{fig:CSeventhor}
\end{center}
\end{figure}
\fig{fig:CSeventhor} shows a plot of the event horizon. As apparent from the plot, the CS event horizon lies outside the global event horizon at $r_+$.  Indeed, at late Poincar\'e times, the event horizon approaches the intersection of the global spacetime boundary $\scri^+$ with the Poincar\'e horizon.

\subsection{Event horizon area \& field theory entropy}

One of the most important and physically interesting attributes of the
event horizon is the area of its cross-sections, which we would
usually take to give the entropy of the corresponding field theory state. We have
seen that the horizon is dynamical, so we expect the area to be
varying. Although the induced metric on the horizon \req{hormet} does not show explicit time dependence, the area variation arises because of the caustics; if a cross-section of
the horizon intersects the line of caustics, it will be parametrized
by $\ell$ lying in some range $0 \leq \ell \leq \ell_{{\rm max}}$,
where $\ell_{{\rm max}}$ is determined by the intersection of the
cross-section with the caustic line. Since the induced metric on the
cross-section is given by \eqref{hormet}, it will then have area 
\begin{equation}
\CA = 2\, \int_0^{\ell_{{\rm max}}} \, \frac{d\ell}{1-\ell^2} = 2 \,
    {\rm arctanh} \,\ell_{{\rm max}}. 
\label{areagen}
\end{equation}	
Note that $\ell_{\max} > \tanh(r_+ \pi)$, so $\CA > 2\pi r_+$, that is,
the area of the dynamical event horizon on the CS spacetime is always
greater than the area of the static horizon in the global BTZ
spacetime, as we would expect. 

A simple family of cross-sections to consider in the global
 coordinates is the intersection with surfaces of constant
 $r$. From \eqref{btzcaus}, we see that for these cross-sections
 $\ell_{{\max}}$ is given by
\begin{equation}
\ell_{{\max}}  = \frac{r \sinh (\pi r_+)}{\sqrt{r_+^2 +r^2 \, \sinh^2(\pi r_+)}},
\label{btzlmax}
\end{equation}	
so the area is
\begin{equation}
\CA =2 \, {\rm arctanh}\left(\frac{r \sinh (\pi r_+)}{\sqrt{r_+^2 +r^2 \, \sinh^2(\pi r_+)}}\right),
\label{arear}
\end{equation}	
which grows logarithmically at large $r$. 

To translate the variation of the area into a statement of the
boundary field theory entropy, we could consider pulling back the area
element of the spatial sections of the horizon along radially ingoing
null geodesics, as advocated in \cite{Bhattacharyya:2008xc}. This
would relate the cross-sections at constant $r$ to some set of
spacelike slices in the field theory on Minkowski space. However, it
seems more natural to look at slices of constant Poincar\'e time $t$,
as this is the natural time coordinate from the field theory point of
view. We would therefore like to check that the area of cross-sections
of constant Poincar\'e time exhibits a similar behaviour.

\begin{figure}[t]
\begin{center}
\includegraphics[scale=0.9]{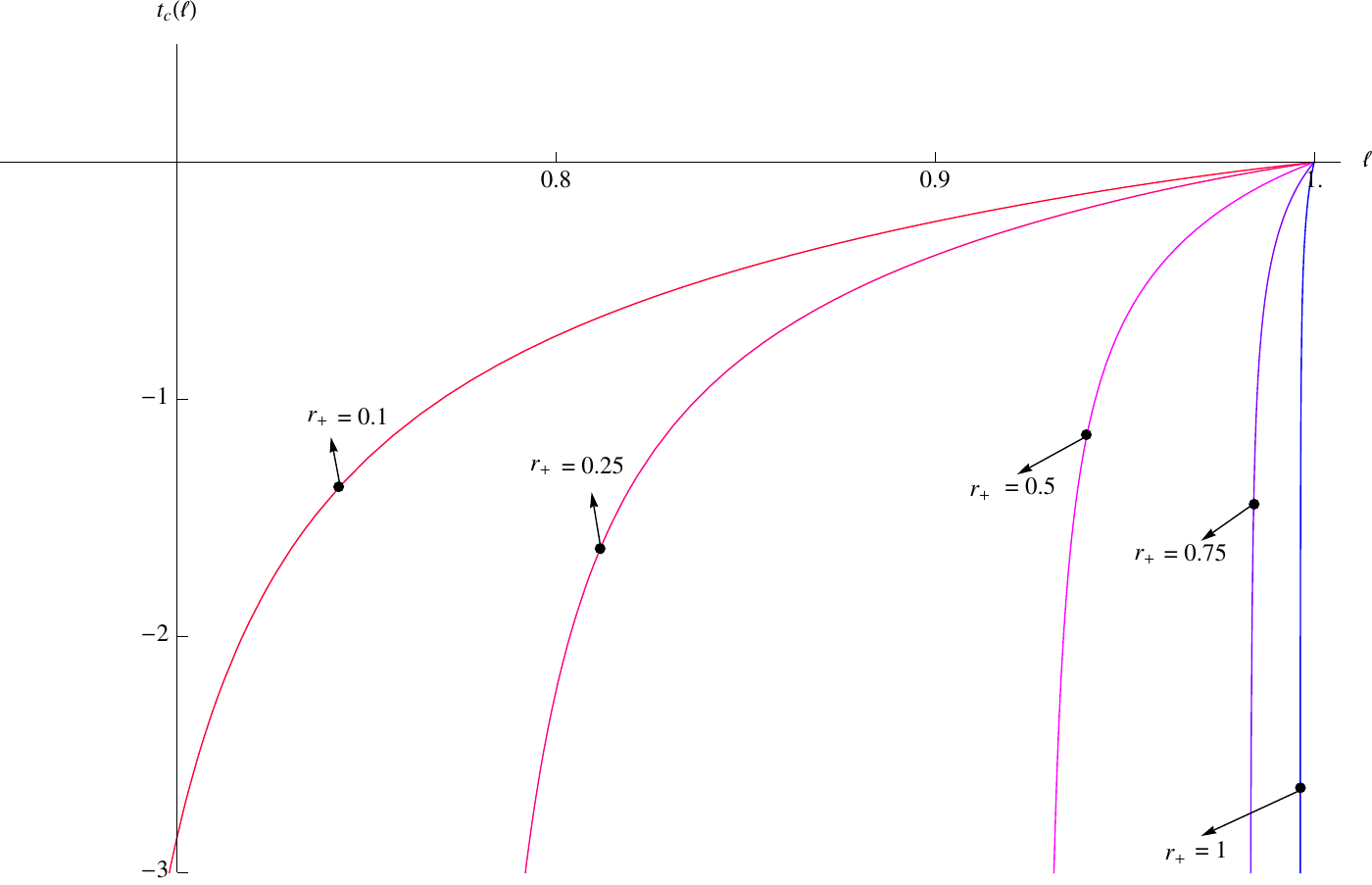}
\caption{Behaviour of the Poincar\'e time along the line of caustics $t_c(\ell)$. We have plotted here the situation for a sampling of BTZ horizon size. As explained in the text, as $r_+ > 1$ (recall that we normalize $L_{\text{AdS}} = 1$) the allowed domain in $\ell$ shrinks exponentially.}
\label{fg:tcvsl}
\end{center}
\end{figure}

We have parameterized the event horizon by $\tau(r,\ell)$ and
$\phi(r,\ell)$ in \req{gBTZrl}. In addition the slices of constant
Poincar\'e time are given explicitly in the second equation of
\req{gtop}. These three relations can be solved to find the desired
cross-sections of the horizon. In particular, along the line of
caustics, the Poincar\'e time is given by 
\begin{equation}
t_c(\ell) = \frac{ \sqrt{r_c^2(\ell)+1} \,  \sin \tau_c(\ell)}{\sqrt{r_c^2(\ell) +1}\,\cos{\tau_c(\ell)}
  - r_c(\ell)},
  \label{tcl}
\end{equation}
where $r_c(\ell)$ and $\tau_c(\ell)$ are given in \eqref{btzcaus}. We
can then determine $\ell_{\max}$ on a cross-section of the horizon at
constant Poincar\'e time by choosing $t_c$ and solving this equation for
$\ell$. Since it is a complicated transcendental expression, it will
not be possible to solve it analytically. 

However, we can make some general remarks.  The slices $t= 0$ in
Poincar\'e coordinates and $\tau = 0$ in global coordinates
coincide. Hence, the $t=0$ cross-section of the horizon is the same as
the $\tau=0$ cross-section, and hence has $\ell_{\max} = 1$. Further,
the slices of $t> 0$ ($<0$) lie entirely in the region $\tau >0$
($<0$) in the global coordinates, while the curve of caustics lies
only in the region $\tau < 0$. Hence for any slice with $t \geq 0$,
$\ell_{\max} = 1$. All of these slices hence have the same
logarithmically divergent area. For the slices with $t <0$, the area
increases monotonically, diverging as $t \to 0$.  Thus, in this
slicing as well, we see a logarithmic divergence of the area; the
potentially surprising feature is that this divergence occurs at $t=0$
in this slicing.\footnote{
Although this observation appears to imply that $t=0$ is special, which is 
rather surprising given that the CFT state evolves smoothly through $t=0$, 
this is really an artifact of our slicing \req{gtop}.  Albeit natural, the constant $t$ slices of \req{gtop} are by no means unique; had we picked the bulk slice anchored at $t=0$ on the boundary to pass through negative $\tau$ in the bulk, we would see the area divergence at a later $t$.
}

Furthermore, $\ell_{\text{max}}$ is constrained to be greater than $\ell_{\text{cmin}}$ determined earlier in \req{lmin}. This is because the Poincar\'e slice extends only down to $t \to -\infty$ (the past Poincar\'e edge) and not all the way down to $\tau = -\infty$. In particular, this implies that even for $t < 0$ the area of the CS event horizon, whilst finite, is nevertheless larger than the area of the event horizon for the global \SAdS{} black hole.

Thus, we have seen that the area of the cross-sections of the global
event horizon in the CS spacetime increases with time, with a
logarithmic divergence. This implies that we cannot identify it with
the entropy in the dual field theory. In the field theory, passing
from the global BTZ black hole to the CS spacetime is just a conformal
transformation, and the entropy of the fluid is invariant under
conformal transformations. Thus, we expect the total entropy of the
fluid flow corresponding to the CS spacetime to be the same as in the
static fluid dual to the global BTZ black hole, independent of the
spatial slice on the boundary we choose to measure it on. The point is
that although the dual fluid flow is time-dependent, it is an ideal
fluid, and in the absence of any viscous or dissipative effects we
cannot have any entropy production (for a brief review see
\App{cshydro}). As a result we should have the field theory entropy
being constant even in the CS spacetime.

We are thus led to propose that in dynamical spacetimes with
`significant' time variations, one should not associate the area of the
event horizon to the entropy of the dual field theory. This point of
view seems natural from studies of entanglement entropy in AdS/CFT
\cite{Hubeny:2007xt} and also in dynamical black hole spacetimes which
are out of the hydrodynamic regime \cite{Chesler:2008hg}. The physical
argument is simply that the event horizon is a teleological object. We
need to know the entire future evolution of the classical geometry in
order to determine the location of the horizon. 
On the other hand, even in a system perturbed away from equilibrium, 
one expects that entropy is
produced locally, i.e., it makes sense to extract the entropy in some
domain of the fluid by analyzing the local evolution equations. One
should not have to evolve the fluid globally for all times before
inferring the entropy production in some region. This suggests that in
the gravitational description one should look for an appropriate
quasi-local horizon whose area we can associate with the entropy. We
will argue that in the CS spacetime the relevant object is the apparent horizon.

This might appear to contradict the argument of
\cite{Bhattacharyya:2008xc} where it was proposed that it is the event
horizon area that corresponds to the field theory entropy. However, as
discussed in \sec{intro}, in that case it was assumed that the fluid
settles down at late times to a stationary solution, which is not the
case for the CS spacetime. Indeed if the dissipative physics drives
the evolution then we expect that at late times the event horizon
coincides with the apparent horizon,\footnote{This statement relies on
a sensible choice of foliation of the spacetime, as the apparent
horizons are foliation dependent. We return to this issue in the
Discussion section and \App{KillingAH}.} and moreover for slow variations which are
required for the hydrodynamic description this situation will pertain
for all times. However, the CS spacetime doesn't fit into the slow
variation paradigm despite the ideal fluid description and hence leads
to a distinct behaviour of event and apparent horizons.

\subsection{Apparent horizon for the CS spacetime}
\label{csah}

We have seen that the event horizon in the CS geometry deviates 
significantly from the global event horizon, 
because of the restriction to $\scri^+_{CS}$. We would now
like to see where the apparent horizon in the CS geometry lies. As the
apparent horizon is a more local concept, we might expect it to be
less affected by the boundary restriction, and this is indeed what we find. 

The notion of apparent horizon is intimately tied to the notion of
trapped surfaces.  Recall that a closed, co-dimension two spacelike
surface $\CS$ (which for the CS$_{2+1}$ geometry is just a closed
curve) is trapped if both (ingoing and outgoing) future-directed null
geodesic congruences emanating normal to the surface $\CS$ have negative
expansions, i.e.\ the areas of the `wavefronts' for these null
congruences decrease in time.  Physically, the presence of such a
trapped surface indicates a region of strong gravitational effects,
since ordinarily, e.g.\ in flat spacetime or AdS, the ingoing
congruence contracts but the outgoing congruence expands.  Indeed, for
spacetimes with complete scri satisfying certain positive energy
conditions, any trapped surface must be contained within a black hole.
Moreover, the existence of a trapped surface implies the existence of
a spacetime singularity.
A surface is marginally trapped if the outgoing null congruence has
zero expansion, while the ingoing congruence has negative expansion.
There are several (strictly-speaking distinct) notions of ``apparent
horizon".\footnote{An excellent review of the various quasi-local
  horizons is found in \cite{Booth:2005qc}.}  In the numerical relativity community,
an apparent horizon on a given spacelike slice is defined as the
outermost marginally trapped surface on that slice.  In mathematical
relativity, an apparent horizon is usually taken to be the boundary of
the union of all trapped points (points lying on a trapped surface),
again on a given spacelike slice.  However, subject to certain smoothness
conditions, which are satisfied by our CS spacetime, the apparent horizon so defined does indeed have vanishing
outgoing expansion \cite{Wald:1984ai}.  

The important subtlety to note about both of these definitions is that
a given spacetime geometry does not by itself specify the location of
the apparent horizon; we first need to specify a foliation of the
spacetime, with respect to which we can then define the apparent
horizon.\footnote{In fact, as demonstrated in \cite{Wald:1991vh}, even the
  Schwarzschild black hole spacetime admits (sufficiently bizarre)
  foliations for which there are no trapped surfaces at all, so that
  there is no apparent horizon.}  In the present case, the physically
relevant foliation is one corresponding to constant Poincar\'e time
slices.  

We could now proceed to find the apparent horizon by an explicit
computation (see \App{csapphor}).  However, this is a difficult
calculation, so it is better to argue on general grounds. Since the
\SAdS{} spacetime satisfies the energy conditions and has a complete
$\scri^+$, the apparent horizon on any spacelike slice must lie inside or on
the event horizon.  The position of the apparent horizon on any given slice
does not depend on the rest of the foliation, so we can view a given
constant $t$ slice as part of a $t$-foliation of the CS, or as part of
a foliation of the global BTZ spacetime (obtained by translating the
given slice by $\tau$ rather than $t$).  In the latter case, the
relevant event horizon is not the CS event horizon which `flares out',
but rather the global event horizon, which stays at constant radius $r
= r_+$ for all times.  This means that the apparent horizon cannot lie
outside the $r = r_+$ surface. Since the $t=0$ slice in Poincar\'e
coordinates coincides with the $\tau=0$ slice in BTZ coordinates, where the apparent horizon of the static black hole coincides with its event horizon, we
know the apparent horizon on this slice coincides with the global
horizon at $r = r_+$.  Furthermore, by the area theorem pertaining to the apparent horizon, this apparent horizon cannot recede in the future since its area cannot decrease.  This, combined with the previous argument that the apparent horizon cannot lie outside the global event horizon, forces the apparent horizon to coincide with the global event horizon for all times, which immediately implies that  its area is constant.\footnote{
In fact, below and in \App{KillingAH} we will argue more generally that an apparent horizon of a spacetime with compact Killing horizon must coincide with this Killing horizon for any foliation which allows complete sections of the Killing horizon.
} The above argument is confirmed by an explicit calculation of the expansion of the null normals in \App{csapphor}.

Since the area of the apparent horizon is constant, it can be
identified with the entropy of the dual field theory. This example
thus provides strong evidence that the entropy of the field theory
fluid should in general be identified with the area of the apparent horizon rather than that of the event horizon. The large difference between the
event horizon and the apparent horizon in this spacetime arises from
the global structure -- specifically the restriction to
$\scri^+_{CS}$. This indicates that it is the teleological nature of
the event horizon which makes it inappropriate for a dual description
of the field theory entropy. The apparent horizon, like the entropy,
is determined by considering the situation at a moment in time (on a
single spacelike slice). Furthermore, subject to the appropriate energy conditions being satisfied the apparent horizon also respects the second law, i.e., the area along cross-sectional slices of the apparent horizon is constrained to be non-decreasing. Hence, it is appropriate to use the pull-back\footnote{Although in \cite{Bhattacharyya:2008xc} it was convenient to pull-back the horizon area form along ingoing null geodesics, in the present case of the apparent horizon being static, it doesn't matter exactly how we pull back the area form to the boundary.} of the area of the apparent horizon to the boundary and regard it as a Boltzmann H-function.

\section{Discussion}
\label{disc}

In this paper we have discussed two distinct hydrodynamic solutions
from a gravitational viewpoint. We analyzed the global causal
properties of the spacetimes dual to Bjorken flow and the conformal
soliton flow. Our bulk analysis confirms that both these spacetimes
fit into the fluid-gravity paradigm, albeit in a somewhat novel
fashion.

The first one, the boost invariant Bjorken flow, was shown to have a regular event horizon and is a genuine regular black hole spacetime. Our analysis here relied on explicitly constructing the null generators of the event horizon order by order in the late time expansion, consistent with the perturbation expansion of the gravity solution. This provides a final consistency check of the late time expansion as a gradient expansion used in the hydrodynamic context.

Our second example, the conformal soliton flow, which from the bulk
standpoint might seem more prosaic, in fact proved more
interesting. The solution was just a coordinate transformation of a
known static solution, the \SAdS{} black hole. However, here we
encountered a surprising result for the event horizon as seen by an
observer living in the boundary Minkowski space. We constructed the
event horizon by explicitly delineating the boundary of the past of
$\scri^+_{CS}$, the future infinity accessible to such a boundary
observer.  While $\scri^+_{CS}$ is not a complete future null
infinity, it is of relevance in the field theory (or hydrodynamic)
description. We can think of the event horizon thus defined as the
Poincar\'e horizon for the black hole spacetime. It is a dynamical
null hypersurface, whose spatial cross-section area diverges
logarithmically for positive Poincar\'e times. We then argued that for
sensible foliations of the spacetime, including the constant
Poincar\'e time slices, the apparent horizon on the slices will
coincide with the global BTZ event horizon at $r=r_+$.

This example shows that in strongly time-dependent settings, it is the
apparent horizon, not the event horizon, which encodes the field
theory entropy in the gravitational dual. One might have thought that
the event horizon demarcates the region of spacetime that the
asymptotic (or boundary) observer can see and therefore its area
should encode the number of active degrees of freedom that are
relevant for the field theory dynamics and hence is a measure of
entropy.  On the other hand, the event horizon is teleological
as its determination requires knowledge of the entire future evolution
of the spacetime. Thus by using its area as a measure of entropy we
would be predicting a drastic non-locality in the field theory
dynamics. We therefore argue that we should instead use the area of the
apparent horizon as a measure for the entropy of the field theory.

Let us briefly revisit the issue of foliation-dependence of apparent
horizons.  In a case of genuinely dynamical spacetimes where the
apparent horizon is a dynamical (spacelike) horizon, general
foliations which don't respect spherical symmetry generically lead to
distinct apparent horizons.  One might expect this to be the case
even for static black holes, since certain sufficiently bizarre
slicings can remove the apparent horizon entirely (as demonstrated for
the Schwarzschild black hole by \cite{Wald:1991vh}). One might
therefore think that by continuity any non-spherical slicing would
deform the position of the apparent horizon.  However, this is not the
case for static black holes.  Above we have argued that in the
Poincare slicing, the apparent horizon of the CS geometry coincides
with the global event horizon at $r = r_+$; but our argument did not
depend on any specifics of the slicing.  In fact, for {\it any}
foliation which admits a complete slice of the global event horizon,
it is easy to show that the apparent horizon must coincide with the
global event horizon.  This can of course be confirmed by explicit
calculation of the expansion; but a much simpler argument is presented in 
\App{KillingAH}.
 The way that
the example of \cite{Wald:1991vh} gets around this is that their
slicing does not allow a complete slice of the future event horizon.
This result is consistent with our expectations from the field theory: for
time-dependence which is trivial in this sense, the entropy should not
change. 

 On the other hand, for genuinely dynamical situations, where
the geometry has no Killing horizons, one might worry that the
location, and thereby the area, of the apparent horizon is
foliation-dependent.  A-priori, this is not necessarily inconsistent
with the field theory expectations: different
boundary slicings may lead to different entropy because the field theory
state at different times is different.
Nevertheless, this intriguing picture of apparent horizon area giving
the entropy of the boundary state still leaves more to be understood in the
genuinely dynamical context: On the one hand, the area of a particular
slice of the apparent horizon depends on where the slice intersects
the horizon -- for expanding apparent horizon, slices intersecting the
horizon at later times will have larger areas.  On the other hand, in
the boundary theory, the entropy should depend only on the state at a
particular boundary time-slice (not necessarily constant Poincare
time, but at the same time not be dependent on the behaviour of the bulk
slice away from the boundary).  This seems to imply that our bulk
prescription has more freedom or ambiguity in defining the entropy
than that afforded by the boundary theory. 
One possibility is that there is a preferred foliation of the bulk, such as a zero-mean-curvature slicing, on which one is supposed to evaluate the area.  However, we don't have a good physical justification for this option.
A simpler resolution to
this puzzle is that in the regime where the concept of entropy is
meaningful, the horizon has to be evolving slowly enough that there is
negligible difference between the areas of all slices of horizon which
end on the same boundary time-slice.  This is essentially the same
picture as that advocated in
\cite{Bhattacharyya:2008jc,Bhattacharyya:2008xc}, except that here we
use it for apparent horizon rather than the event horizon.
In effect the field theory on the boundary should achieve local equilibrium in order for entropy to be a meaningful observable.

A general lesson from this discussion seems to be that while in
situations near equilibrium, the event and apparent horizons are
reasonably close and therefore provide adequate diagnostic measures
for the field theory entropy, in far-from-equilibrium scenarios or
those with strong modifications to the boundary conditions, we should
use the quasi-local apparent horizon as opposed to the event horizon
to measure the entropy. It would be interesting to establish this for
generic situations in the context of the AdS/CFT correspondence.

\paragraph{Note added in v2:}
We have been rather glib in our terminology, mainly because the context in which we are working is sufficiently mild.  Technically speaking, what we called ``apparent horizon" should really be referred to as ``dynamical horizon" (in case of a spacelike co-dimension 1 surface), or ``isolated horizon" (in case of a null surface).  The BF spacetime exemplifies the former, which is the more generic case, whereas the CS spacetime gives the latter.
We should emphasize that generally an apparent horizon is defined as a co-dimension 2 surface, on a given leaf of foliation, corresponding to the outermost marginally trapped surface or the boundary of trapped points, and as such, the set of apparent horizons on all leaves of the foliation need not form a smooth co-dimension 1 surface in the full spacetime.\footnote{In fact for certain discontinuous cases these two definitions don't necessarily coincide.} This supplies further reason why one can not assert that ``entropy is given by the area of apparent horizon" in full generality, since the entropy is expected to be smoothly varying in time, whereas the area of apparent horizon can jump discontinuously.  We would however view the foliation-dependence to be a more worrying issue, as explained in the Discussion above, since it arises under much milder conditions than the actual discontinuities in apparent horizon. 

\subsection*{Acknowledgements}
\label{acks}

It is a pleasure to thank Roberto Emparan,  David Garfinkle, Carsten Gundlach, Michal Heller, Gary Horowitz, Kasper Peeters, Larry Yaffe and Marija Zamaklar for useful discussions.   VH and MR would like to thank  LPTHE (Paris), MIT, and KITP for hospitality during the course of this work. MR would also like to acknowledge the hospitality of CERN.  PF is supported STFC Rolling Grant. VEH, MR and SFR are supported in part by STFC. VEH and MR are supported in part by the National Science Foundation under the Grant No. NSF PHY05-51164.

\appendix

\section{Conformal solitons in higher dimensionss} 
\label{cshighd}

Consider the global \SAdS{d+1} black hole whose metric is given as 
\begin{equation}
 ds^2=-f(r)\,d\tau^2+\frac{dr^2}{f(r)}+r^2\,d\Omega^2_{d-1}\,, 
 \label{eqn:globSchw}
\end{equation}
with the function $f(r)$ being
\begin{equation}
 f(r)=1+r^2-\left(\frac{r_+}{r}\right)^{d-2} (1+r_+^2)
\label{}
\end{equation}	
We choose to parameterize the metric on the $\Sp^{d-1}$ keeping manifest $SO(d-1)$ rotational isometry, i.e., 
\begin{equation}
 d\Omega^2_{d-1}=d\phi^2+\sin^2\phi\; d\Omega_{d-2}^2 \,.
\end{equation}
This makes it easier to make contact with the discussion in \sec{confsol} for the BTZ spacetime, since now \fig{fig:AdScoords} illustrates the behaviour in the $\{r,\tau,\phi\}$ space (where every point now represents a $\Sp^{d-2}$ of radius $r\,\sin\phi$). The event horizon of this spacetime is at $r= r_+$.

To restrict consideration to the Poincar\'e patch of this spacetime, we proceed as before by  a similar coordinate transformation  to \req{gtop}:
\begin{equation}
z = \frac{1}{\sqrt{r^2 +1} \, \cos \tau + r\, \cos \phi} \ , \ \
t = \frac{\sqrt{r^2 +1} \, \sin \tau }{\sqrt{r^2 +1} \, \cos \tau + r\, \cos \phi} \ , \ \
{\bf x}_{d-1} = \frac{ r \, \sin \phi\, {\bf \Omega}_{d-2}}{\sqrt{r^2 +1} \, \cos \tau + r\, \cos \phi} \ .
\label{gtopA}
\end{equation}	
with ${\bf \Omega}_{d-2}$ denoting a unit vector on $\Sp^{d-2}$. 
The resulting CS$_{d+1}$ spacetime is qualitatively similar to that discussed in \sec{confsol}.  We now wish to determine the event horizon for this CS$_{d+1}$ spacetime.
This is achieved as discussed in \sec{cseh} by working out the null geodesics bounding the past of $\scri^+_{CS}$.  

Taking into account the symmetries of the background \eqref{eqn:globSchw}, the equations for a null geodesic congruence respecting the $SO(d-1)$ rotational symmetric are
\begin{equation}
 \dot r^2=1-\frac{\ell^2}{r^2}\,f(r)\,,\qquad \dot\tau=\frac{1}{f(r)}\,,\qquad \dot\phi=\frac{\ell}{r^2}\,, \label{eqn:nullgeods}
\end{equation}
where $\dot\,$ denotes the derivative with respect to the affine parameter $\lambda$. We have normalized the affine parameter by choosing to set the energy of the geodesic to unity, and we identify each geodesic by the parameter $\ell$ corresponding to the angular momentum.  

We are interested in the null geodesics that can reach $\scri^+$ when
sent from some $r>r_+$.  The only novelty in this calculation relative
to the BTZ case is that the effective potential for the radial motion
(writing the geodesic equation as $\dot r^2 + V_{\text{eff}}(r) = 0$),
\begin{equation}
V_{\text{eff}}( r) = -1+ \frac{\ell^2}{r^2}\,f(r)\ ,
\label{}
\end{equation}	
has a distinct maximum associated with the unstable photon orbit at 
\begin{equation}
r_{\text{ph}} = r_+ \, \left(\frac{d}{2}\, (1+r_+^2)\right)^{\frac{1}{d-2}}.
\label{}
\end{equation}	

Geodesics emanating from $r < r_{\text{ph}}$ will make it out to the boundary only if 
$V_{\text{eff}}(r_{\text{ph}}) < 0$, which translates to an upper bound on the angular momentum $\ell < \ell_{\text{max}}$
\begin{equation}
\ell^2_{\text{max}}  = \frac{r_{\text{ph}}^2}{r_{\text{ph}}^2+\left(1- \frac{2}{d}\right) }.
\label{}
\end{equation}	
For $r > r_{\text{ph}}$ however we can have $0 \le \ell \le 1$ as usual. Massless particles have to overcome the gravitational centripetal barrier to escape to infinity to $\scri^+$.

Despite this complication, it is straightforward to integrate \eqref{eqn:nullgeods} to find the geodesics explicitly. For the special case of $d =4$ (i.e., \SAdS{4+1}) one can write closed form expressions using $u = 1/r$ as:\footnote{We have picked the branch cuts in evaluating the integrals so as to obtain manifestly real expressions for $\tau$ and $\phi$ in \req{eqn:nullgeodsres}.}
\begin{equation}
\begin{aligned}
 \phi(u,\ell)&=\pm\frac{1}{\alpha\,r_+\sqrt{1+r_+^2}}
	\,F\left[\textstyle\arcsin\left(\frac{u}{\alpha}\right),\frac{\beta}{\alpha}\right],\\
\tau(u,\ell)&=\pi-\frac{1}{\beta\,\ell\,r_+\,\sqrt{1+r_+^2}(1+2\,r_+^2)}\left\{\frac{1}{\zeta^2}\,\Pi\left[\textstyle \arcsin\left(\frac{u}{\alpha}\right),-\frac{\alpha^2}{\zeta^2},\frac{\alpha}{\beta}\right]+\frac{1}{\xi^2}\,\Pi\left[\textstyle \arcsin\left(\frac{u}{\alpha}\right),-\frac{\alpha^2}{\xi^2},\frac{\alpha}{\beta}\right]\right\}\,, \label{eqn:nullgeodsres}
\end{aligned}
\end{equation}
where $F(\varphi,k)$ and $\Pi(\varphi,n,k)$ are the incomplete elliptic integrals of the first and third kind respectively, and we have defined the constants
\begin{equation}
\begin{aligned}
 \alpha=\frac{1}{2\,r_+^2(1+r_+^2)}
	\left(1+\frac{1+2\,r_+^2}{\ell}\sqrt{\ell^2-\ell_{\text{max}}^2}\right)
\,,&\qquad
\beta=\frac{1}{2\,r_+^2(1+r_+^2)}\left(1-\frac{1+2\,r_+^2}{\ell}\sqrt{\ell^2-\ell_{\text{max}}^2}\right)\,,\\
\zeta=\frac{1}{\sqrt{1+r_+^2}}\,,&\qquad \xi=\frac{1}{r_+}\,.
\end{aligned}
\end{equation}
The event horizon determined by this null congruence qualitatively looks similar to the BTZ case as illustrated in \fig{fig:CSeventhor}.

 \section{The conformal soliton flow and hydrodynamics}
 \label{cshydro}
 We illustrate the fact that the conformal soliton flow does not lead to entropy production. As explained in the text, the  coordinate transformation \req{gtop} when restricted to the boundary is a conformal transformation.  In this appendix we review briefly the conformal transformation of the hydrodynamic variables.
  
In $d$ spacetime dimensions, under a conformal transformation of the background metric, the stress tensor transforms homogeneously with conformal weight $d+2$.  In particular, we have the transformation:
\begin{equation}
g_{\mu \nu} = e^{2\phi} \, \tilde{g}_{\mu \nu} \ , \qquad  T^{\mu \nu} = e^{-(d+2)\phi} \, \tilde{T}^{\mu \nu}. 
\label{cmap1}
\end{equation}	
which implies that the velocity and thermodynamic variables transform as 
\begin{equation}
u^\mu = e^{-\phi} \, \tilde{u}^{\mu} \ , \quad \rho = e^{-d\phi} \, \tilde{\rho} \ , \quad P = e^{-d\phi}\, \tilde{P} \ , \quad T = e^{-\phi} \, \tilde{T} \ , \quad s = e^{-(d-1)\phi} \, \tilde{s}
\label{cmap2}
\end{equation}	
where $s$ is the entropy density of the fluid.  Note that the total entropy $S$ is clearly invariant under conformal transformations.\footnote{The entropy is dimensionless and therefore doesn't depend on the conformal frame. Entropy density on the other hand behaves like inverse spatial volume as it must.} One further quantity we will be interested in is the entropy current, which for an ideal fluid takes the form\footnote{For the moment we are going to ignore corrections to this coming from dissipative terms which of course lead to entropy production.} 
\begin{equation}
j^\mu_s = s \, u^\mu ,
\label{}
\end{equation}	
and transforms under conformal transformations as 
\begin{equation}
j^\mu_s = e^{-d \phi} \, \tilde{j}^\mu_s .
\label{cmap3}
\end{equation}	

For simplicity we will take the tilded variables to correspond to the global BTZ solution. We have then in the global coordinates 
\begin{equation}
\tilde{u}^a= \left(\frac{\partial}{\partial \tau}\right)^a \ , \quad \tilde{T} = \frac{r_+}{2\pi} \ , \quad \tilde{s} =  \frac{1}{4} \, r_+  ,
\label{globalqs}
\end{equation}	
leading to an entropy current vector
\begin{equation}
\tilde{j}_s^a = \frac{1}{4} \, r_+ \, \left(\frac{\partial}{\partial \tau}\right)^a \ ,  
\label{globalec}
\end{equation}	
which clearly is divergence free $\widetilde{\nabla}_a \tilde{j}^a_s = 0$.

Transforming to the Poincar\'e coordinates we find the velocity 1-form 
\begin{equation}
u = \frac{1}{2\, W} \, \left( (1+x^2+t^2) \, dt - 2\, t\, x\, dx\right) ,
\label{up1form}
\end{equation}	
which leads to an entropy current
\begin{equation}
j_s^a = \frac{r_+}{8 \, W^2} \,  \left( (1+x^2+t^2) \, \left(\frac{\p}{\p t} \right)^{\! a} - 2\, t\, x\, \left(\frac{\p}{\p x} \right)^{\! a} \right),
\label{}
\end{equation}	
which again turns out to satisfy
\begin{equation}
\nabla_a j^a_s  =0 
\label{}
\end{equation}	
which is what we expect. The system stays an ideal fluid in the Poincar\'e frame. While there is some spatio-temporal variation of the energy density and temperature, there is no entropy production.  This is of course as expected, an ideal fluid stays ideal in all conformal frames.
While we have worked out the result for the BTZ spacetime, it is easy to check that the same result holds in higher dimensions. In fact from the discussion in \App{cshighd} it is clear that the transverse $SO(d-1)$ symmetry of the $\Sp^{d-2}$ ensures that we are essentially dealing with very similar physics.

\section{Apparent horizon in the Poincar\'e slicing of BTZ}
\label{csapphor}

In this Appendix we derive the apparent horizon for the Poincar\'e patch of the BTZ spacetime i.e., the CS$_3$ spacetime. We have given a general argument in \sec{csah} to claim that the apparent horizon for the CS spacetime coincides with the event horizon in the global BTZ spacetime, and in \App{KillingAH} we generalize this still further, to argue that for any stationary black hole, in any foliation admitting a complete section of the horizon, the apparent horizon must coincide with the event horizon. However, to provide more concrete insight, here we proceed by explicit calculation.
 
The coordinate transformation in the bulk \req{gtop} mapping the global BTZ to CS, turns out to be too cumbersome for computation. In order to implement a conformal transformation on the boundary to map the fluid on $\R^{1,1}$ to the cylinder $\Sp^1 \times \R^1$, we only require a bulk coordinate transformation that reduces to the appropriate conformal mapping \req{bdyconf}. As a simpler bulk diffeomorphism consider:
\begin{equation}
\zs = \frac{1}{r \, ( \cos \tau + \cos \phi)} \ , \ \
\ts = \frac{\sin \tau }{\cos \tau + \cos \phi} \ , \ 
\xs = \frac{\sin \phi}{\cos \tau + \cos \phi} \ .
\label{gtopsimpler}
\end{equation}	
We will use  \req{gtopsimpler} and find the apparent horizon of the $t={\it const.}$ slices.\footnote{Strictly speaking the constant $\ts$ slices in the coordinates \req{gtopsimpler} differ from the constant Poincar\'e time slices in \req{gtop}. Nevertheless, the foliations are sufficiently similar that we can trust that the apparent horizons in the two coordinate charts with $\ts = {\it const}$ spacelike slices have the same qualitative features.}

As a further simplification, we will only consider two slices: one at $\ts=0$, which is a symmetric slice and another  at $\ts= \infty$. We will argue that these slices have a marginally trapped surface at $r = r_+$ which will be the apparent horizon we seek. Since the location of the apparent horizon is the same on these two distinct slices, using the monotonicity property of  apparent horizon area, we conclude that the apparent horizon must lie at $r=r_+$ for all $\ts$.
Moreover to keep the equations manageable we will write them out in the global coordinates, with the transformation \req{gtopsimpler} being used only to specify the slices.

\paragraph{The time symmetric slice:} The $\ts=0$ slice, $\Sigma_0$,  clearly coincides with the $\tau=0$ slice in global coordinates. On this surface we consider an arbitrary closed curve $\gamma$ given by  $r = g(\phi)$, which is our ansatz for a trapped surface. The outgoing null normal to the spacetime co-dimension two surface $\gamma$ is given as 
\begin{equation}
k_a = -\sqrt{g^2-r_+^2}\,(d\tau)_a +\frac{1}{\sqrt{g^2-r_+^2+\frac{(g')^2}{g^2}}}\big((dr)_a-g'(d\phi)_a\big)
\label{nullnor}
\end{equation}	
Using the induced metric on $\gamma$:
\begin{equation}
 q_{ab}\,dx^a\,dx^b=\left[g^2+\frac{(g')^2}{g^2-r_+^2}\right]\,d\phi^2\,,
\label{gameq}
\end{equation}
the expansion of the outgoing null normals $\theta \equiv q^{ab}\nabla_a k_b$ can be computed to be  
\begin{equation}
 \theta_0  =\frac{g^2(g^2-r_+^2)^2}{\big[g^2(g^2-r_+^2)+(g')^2\big]^{\frac{5}{2}}}
	\,\big[(g')^2+g^2(g^2-r_+^2)-g\,g''\big]\,. \label{expansion}
\end{equation}
We want to find a marginally trapped surface for which the outgoing null geodesics are non-expanding. This requires us to have $\theta_0 = 0$ which gives us a second order non-linear ordinary differential equation  for the curve $r = g(\phi)$. We will now argue that in fact the curve is simply $g(\phi)= r_+$, which coincides with the location of the global event horizon.

Along any curve $\gamma$, consider the outermost point $p$, where the function $g(\phi)$ attains its maximal value.  Then $g$ will satisfy $g'(p)=0$ and $g''(p)<0$, which allows us to bound the expansion of the outgoing null geodesics at this point:\begin{equation}
 \theta_\gamma(p)=\frac{\sqrt{g^2-r_+^2}}{g}\left[1-\frac{g''}{g(g^2-r_+^2)}\right]
	\bigg\vert_p\geq0\,, \qquad \textrm{for}\qquad g>r_+>0\,.
\end{equation}
This implies that if $\gamma$ is a trapped surface, then its furthest point cannot lie outside $r_+$.  
On the other hand,  if we consider a circle $\mathcal C$, for which $g'=g''=0$, then we find that for all points on $\mathcal C$, the expansion of the outgoing null geodesics is given by
\begin{equation}
 \theta_{\mathcal C}=\frac{\sqrt{g^2-r_+^2}}{g}\,,
\end{equation}
which is always positive if $g>r_+$ and becomes zero when  $g=r_+$.
(Note that if we take $\mathcal C$ to enclose the whole of $\gamma$ and intersect it at $p$ outside $r_+$, then $\theta_{\mathcal C}\leq \theta_\gamma(p)$,  since $\gamma$ is more curved than $\CC$  at $p$.)
We conclude then that, since no trapped surface can reach outside $r=r_+$, whereas the circle $g=r_+$  is marginally trapped, the curve $g(\phi)=r_+$ gives the apparent horizon. Therefore, we have argued that in the $t=0$ slice of the conformal soliton geometry, the apparent horizon coincides with the global event horizon. 

\paragraph{The late Poincar\'e time slice:} From \eqref{gtopsimpler}, we find that the $t=\infty$ slice, $\Sigma_\infty$, is given by the condition
\begin{equation} 
 \cos\tau+\cos\phi=0.
\end{equation}
Following the same steps as before, we consider an arbitrary closed curve on $\Sigma_\infty$ and compute the expansion of the outgoing null geodesics to this curve. We find
\begin{equation}
\theta_\infty=\frac{r_+(g^2-r_+^2)^2}{\big[r_+^2(g^2-r_+^2)+(g')^2\big]^{\frac{5}{2}}}
	\,\big[g\,(g')^2+g\,r_+^2(g^2-r_+^2)-g_+^2\,g''\big]\,.
\end{equation}
Applying the same argument as in the previous case we conclude that the apparent horizon of the $\Sigma_\infty$ slice is at $g=r_+$, which again coincides with location of the global event horizon. Therefore, since the area of the apparent horizon does not change with time, we conclude that the entropy in the field theory also stays constant, as expected. 

\paragraph{General slicings:} We have given in \sec{csah} a general argument for the apparent horizon to coincide with the global event horizon. Here we illustrate this by an explicit computation.  To avoid obfuscating issues to do with foliation dependence of the apparent horizon, we will focus on slices which contain an entire spatial cross section of the global event horizon.

Consider a general timelike foliation of the \SAdS{} spacetime given by an arbitrary function
\begin{equation}
t_g = F(\tau, r, \phi) \ . 
\label{gent}
\end{equation}	
We are interested in the nature of trapped surfaces lying on the spacelike surfaces defined by \req{gent}. We find it convenient to invert the relation \req{gent} and express the global time coordinate $\tau$ as a function of the other variables; 
\begin{equation}
\tau = \CF(t_g;r,\phi)\,,\label{eqn:generalslicing} \ .
\end{equation}

On each of the slices \eqref{eqn:generalslicing} consider the circles $r=$ const. The outgoing null normal to any of these circles is given by
\begin{equation}
 k_a=-\mathcal N(d\tau)_a+\left(\mathcal N\,\partial_r \CF+\frac{r}{\mathcal N\, Q}\right)(dr)_a+\mathcal N\,\partial_\phi \CF\,(d\phi)_a\,,
\end{equation}
where
\begin{equation}
\mathcal N=\sqrt{\frac{r^2(r^2-r_+^2)}{r^2-(r^2-r_+^2)\left[r^2(r^2-r_+^2)(\partial_r \CF)^2+(\partial_\phi \CF)^2\right]}}\;,\qquad
Q=\sqrt{r^2-(r^2-r_+^2)(\partial_\phi \CF)^2}\;.
\end{equation}

While the general expression is unilluminating, for our purposes it suffices to argue that the surfaces $r = r_+$ are trapped. In order to establish this, we compute the expansion of these outgoing null normals for our test circles lying in the vicinity of the global event horizon,  i.e., $r \sim  r_+$. Assuming furthermore that $\CF$ and its derivatives are sufficiently smooth at $r=r_+$, we find
\begin{equation}
 \theta=\frac{\sqrt 2\left[r_++\partial_\phi^2 \CF(r_+,\phi)\right]\sqrt{r-r_+}}{r_+^{\frac{3}{2}}}+\mathcal O\left((r-r_+)^{\frac{3}{2}}\right)\;
\end{equation}
near $r=r_+$. Therefore,  we conclude that the surface $r=r_+$ is indeed trapped for general slicings of the spacetime.

\section{Apparent horizon coincides with Killing horizon}
\label{KillingAH}

Recall that for general dynamical black hole spacetimes, the location of an apparent horizon is foliation dependent.  Changing the foliation slightly will in general change the location of the apparent (or the so-called dynamical) horizon slightly.  However, this is not the case for stationary black holes, or more generally black holes which have a Killing horizon.  (Note that for  stationary black holes, the event horizon is a Killing horizon.)
In this Appendix we will explain why an apparent horizon of a spacetime with compact Killing horizon must coincide with the Killing horizon for {\it any} foliation which allows complete sections of the Killing horizon.

The basic outline of the argument is the following:
Any slice of a Killing horizon is a marginally trapped surface, since the outgoing null
normals to any such slice of the horizon coincide with the horizon
generators (due to the Killing horizon being null), and the horizon
generators have zero expansion (because any spacelike slice of a
Killing horizon has the same proper area).  Moreover, this marginally
trapped surface is the outermost one, since there cannot be trapped
surfaces outside the event horizon.  Hence for a slicing which admits
a complete cross-section of the (future) event horizon, the apparent
horizon necessarily coincides with the event horizon.  

Let us now demonstrate the assertion that any complete slice of a Killing horizon is a marginally trapped surface using the example of 3-dimensional rotationally-invariant black hole.  We proceed by first describing an arbitrary spacelike slice of the horizon in terms of its tangent vector, and then determining the null normals to this vector.
Having obtained the null normals, we can then easily confirm that the outgoing null normal coincides with the horizon generators.

Since we wish to consider the geometry at the event horizon, let us write the metric more conveniently in ingoing Eddington coordinates:
\begin{equation}
ds^2 = g_{vv} \, dv^2 + 2 \, dv \, dr + g_{xx} \, dx^2
\label{gen3}
\end{equation}	
where the metric components $g_{vv}$ and $g_{xx}$ are functions of $r$ which we don't need to specify for our argument.
Suppose the event horizon lies on a constant $r$ surface, $r = r_+$, where $g_{vv}=0$.  Then along any spacelike slice of the horizon, we can write the tangent vector as
\begin{equation}
s^a = \CN \left(\frac{\p}{\p x} \right)^{\! a}  + \CC \, \left(\frac{\p}{\p v} \right)^{\! a} 
\label{tangent}
\end{equation}	
for some arbitrary coefficient $\CC$ (which can vary along the slice).  In order for $s^a$ to be unit-normalised,  it suffices to let $\CN = 1/\sqrt{g_{xx}(r=r_+)}$.
Now, to solve for the null normal to our slice, we want to find a vector $\xi^a$ which satisfies $\xi^a \, \xi_a = 0$ and $\xi^a \, s_a = 0$.
Let
\begin{equation}
\xi^a = \left(\frac{\p}{\p v} \right)^{\! a}  + \CA \, \left(\frac{\p}{\p r} \right)^{\! a}  + \CB \, \left(\frac{\p}{\p t} \right)^{\! a} 
\label{normal}
\end{equation}	
Then the null condition $\xi^a \, \xi_a = 0$ implies
$g_{vv} + 2 \, \CA + \CB^2 \, g_{xx} \mid_{r=r_+}
 = 2 \, \CA + \CB^2/\CN^2 = 0$, 
 while the normal condition $\xi^a \, s_a = 0$ yields
 $\CC \, g_{vv} + \CC \, \CA + \CN \, \CB  \, g_{xx} \mid_{r=r_+} 
 =  \CC \, \CA + \CB /\CN = 0$.
 For any given $\CC$, there exist two distinct solutions: either 
\begin{equation}
\CA = \CB = 0 \ ,
\label{outgoing}
\end{equation}	
or 
\begin{equation}
\CA = - 2 / \CC^2 \ , \qquad \CB = 2 \, \CN / \CC \ .
\label{ingoing}
\end{equation}	
Since in the latter solution, \req{ingoing}, the coefficient $\CA$ of $\left(\frac{\p}{\p r} \right)^{\! a} $ is negative, the resulting $\xi^a$ corresponds to ingoing null normals.  This means that the first solution, \req{outgoing}, corresponds to the outgoing null normals.  Thus we have found that for an arbitrary slice of the horizon, i.e.\ for any $\CA$, the outgoing null normal is given by 
$\xi^a = \left(\frac{\p}{\p v} \right)^{\! a} $, independently of $\CA$.
It is easy to see that if $\left(\frac{\p}{\p v} \right)^{\! a}$ is a Killing vector, the Killing horizon generators are simply $\left(\frac{\p}{\p v} \right)^{\! a}$, which are null on the horizon.  This proves our first assertion, that the outgoing null normals to any slice of the Killing horizon coincide with the horizon generators.

Finally, the fact that the Killing horizon generators have zero expansion can be easily shown by noting that the proper area of the horizon remains constant along the generators, and moreover using similar arguments as above, this area is the same along any spacelike slice of the horizon; we leave this as an exercise for the reader.


\providecommand{\href}[2]{#2}\begingroup\raggedright\endgroup

\end{document}